# Convolutional Neural Network for Behavioral Modeling and Predistortion of Wideband Power Amplifiers

Xin Hu, *Senior Member, IEEE*, Zhijun Liu, Xiaofei Yu, Yulong Zhao, Wenhua Chen, *Senior Member, IEEE*, Biao Hu, Xuekun Du, Xiang Li, Mohamed Helaoui, *Senior Member, IEEE*, Weidong Wang, and Fadhel M. Ghannouchi, *Fellow, IEEE*

***Abstract*—Power amplifier (PA) models, such as the neural network (NN) models and the multilayer NN models, have the problems of high complexity. In this paper, we firstly propose a novel behavior model for wideband PAs using a real-valued time-delay convolutional neural network (RVTDCNN). The input data of the model are sorted and arranged as the graph composed of the in-phase and quadrature (I/Q) components and envelope-dependent terms of current and past signals. We design a pre-designed filter using the convolutional layer to extract the basis functions required for the PA forward or reverse modeling. Then, the generated rich basis functions are modeled using a simple fully connected layer. Because of the weight sharing characteristics of the convolutional model's structure, the strong memory effect does not lead to a significant increase in the complexity of the model. Meanwhile, the extraction effect of the pre-designed filter also reduces the training complexity of the model. The experimental results show that the performance of the RVTDCNN model is almost the same as the NN models and the multilayer NN models. Meanwhile, compared with the models mentioned above, the coefficient number and computational complexity of the RVTDCNN model are significantly reduced. This advantage is noticeable when the memory effects of PA are increased by using wider signal bandwidths.**

## I. Introduction

As an indispensable component in the wireless communication system, power amplifiers (PAs) provide enough power for transmitting the signal through the channel to ensure that the receiver can collect the signal with relatively good signal to noise ratio [1]–[4]. However, PA' nonlinearity and memory effect can lead to the spectral expansion and decrease in Adjacent channel power ratio (ACPR) performance, thereby degrading the quality of communication [5]–[9]. Behavior modeling provides an effective method for nonlinear analysis and modeling of PAs. The behavior modeling often constructs the mathematical nonlinear modeling function by capturing the input and output responses of the system when driven with highly time varying signals to trig and observe the static nonlinear behavior of the system as well as the dynamics of the system are often designated as memory effects [10], [11]. With the incoming 5G standard calling for a sharp increase in data transmission rate up to multiple Gbps, the signal bandwidth needs to be increased significantly up to several hundred MHz. Accordingly, ultra-broadband PA behavior modeling and digital predistortion (DPD) have become the current research hotspot.

Traditional behavioral models with memory effect, including the Volterra model and several compact Volterra models, have been widely used in the modeling of wideband PA [12], [13]. However, the high correlation between polynomial bases in these models makes it difficult to improve the modeling performance [14]. Recently, the outstanding achievements of artificial neural networks (ANNs) in the field of communication have attracted the attention of researchers in the field of wireless PA modeling. Due to ANN's excellent performance for the approximation of nonlinear function, many works published in open literature have studied its application in PA modeling and predistortion area [15]–[19]. When PA exhibits complicated nonlinear characteristics and memory effects, it is difficult to achieve good modeling performance with low-complexity- based ANN models. This motivated this work to address the problem of how to derive broadband low-complexity NN based models that can provide accurate modeling performance for the forward and inverse (predistorters) models.

To address the above issues and inspired by the emergence of artificial intelligence (AI) in the broadband communications area, advanced NN based models will be investigated [20]–[24]. Deep learning [25]–[28] in the AI field has shown excellent performance in discovering complex non-linear relationships using labeled data. In particular, convolutional neural networks (CNNs) [29], [30] and recurrent neural networks (RNNs) [31], [32] in deep learning have been proven to be effective in many fields including wireless communication [33]–[38]. However, according to the results of our research, the work on the use of deep learning to solve the problems of behavior modeling and

This work is supported by the National Natural Science Foundation of China (No. 61701033).

X. Hu (e-mail: huxin2016@bupt.edu.cn), Z. J. Liu (e-mail: lzj2017110489@bupt.edu.cn), X. F. Yu, and W. D. Wang are with the School of Electronic Engineering, Beijing University of Posts and Telecommunications, Beijing 100876, China, and also with the Key Laboratory of Universal Wireless Communications, Ministry of Education, Beijing University of Posts and Telecommunications, Beijing 100876, China.

Y. L. Zhao, X. K. Du, X. Li, M. Helaoui, and F. M. Ghannouchi are with the Intelligent RF Radio Laboratory, Department of Electrical and Computer Engineering, Schulish School of Engineering, University of Calgary, Calgary, AB T2N 1N4, Canada (e-mail: fadhel.ghannouchi@ucalgary.ca).

W. H. Chen is with the Department of Electronic Engineering, Tsinghua University, Beijing 100084, China.

B. Hu is with School of Electronic Science and Engineering, University of Electronic Science and Technology of China, Chengdu 610054, China.

linearization of PAs [14], [39]–[42] is limited. One of the important reasons is that the regression algorithm based on RNNs learning is often utilized for natural speech processing and time series processing tasks. If they are used for modeling and linearization of PAs, although they have fewer parameters compared with feedforward neural networks due to their characteristics of weight sharing, the complex training algorithm seems to make the method complicated [43]. In addition, CNN is usually used as a classifier, and the output layer makes a discrete decision rather than outputting a continuous signal. However, with the present work, it will demonstrate for the first time that CNN can be adapted and used in the fields of behavior modeling and DPD synthesis of the PAs. The NN model's complexity reduction will mainly result from the characteristic of weight sharing in CNN structures [29], [30]. The aspect of increasing the input dimension without changing network structure has attracted our attention.

We firstly apply CNN to PA modeling and propose a real-valued time-delay convolutional neural network (RVTDCNN) behavior model for wideband wireless PA modeling. Due to CNN cannot be directly used to build the PA model since input signals are not graphs, the input data are sorted and arranged as the graph composed of the in-phase and quadrature (I/Q) components and envelope- dependent terms of current and past signals. And then, this model constructs a pre-designed filter using the convolutional layer to extract the basis functions required for PA forward or reverse modeling. Finally, the extracted basis functions are input into a simple fully connected layer to build the PA model. The model complexity of RVTDCNN is significantly reduced due to the weight sharing characteristic of the convolution structure. Meanwhile, the extraction effect of the pre-designed filter also reduces the training complexity of the model. In order to evaluate the model performance of the RVTDCNN model, we compared the RVTDCNN model with other existing models (including NN and multilayer NN model) by experiment and simulation. The results show that, compared with the existing state-of-the-art models, the model performance of the RVTDCNN model, especially the model complexity, is reduced in terms of the number of model's coefficients.

The contributions of this paper are as follows:

- As the signal bandwidth increases, PA exhibits complicated nonlinear characteristics and memory effects. It is difficult to achieve good modeling performance with low complexity-based traditional behavioral models [14]. To address the problem of how to derive the broadband low complexity model, the first CNN-based architecture for extracting the PA behavioral model is proposed to improve nonlinear modeling performance.
- It is found that the existing NN-based models [14], [19] still have a considerable complexity of model coefficients. To alleviate this issue, the input dataset is constructed as the graph and the convolution layer is studied and designed as a pre-designed filter to extract the basis functions required for the PA modeling.
- If RNN or CNN is used in the modeling and linearization of PAs, they have high computational complexity for parameters training [43]. To reduce the computational complexity, a training methodology for PA modeling is proposed to accelerate the training of the PA model.

The remainder of this paper is organized as follows. In Section II, the existing neural network models for PA modeling, including shallow neural network (NN) models and deep neural network (DNN) models, are briefly reviewed. Section III proposes the structure of the RVTDCNN model and describes it in a detailed manner. Section IV discusses the training process of the RVTDCNN model and analyzes the model complexity of the RVTDCNN model. Section V describes the platform for experimental validation. Section VI reports the measurement and validation results and compares the proposed model with other models. Finally, Section VII gives the conclusions.

## II. ANNs for PA Modeling

### A. *Shallow Neural Networks for PA modeling*

The shallow NN networks with fewer hidden layers are used to express the output characteristics of the PA due to its relatively simple network structure and training process, as shown in Fig. 1 (a) and (b) [15], [19].

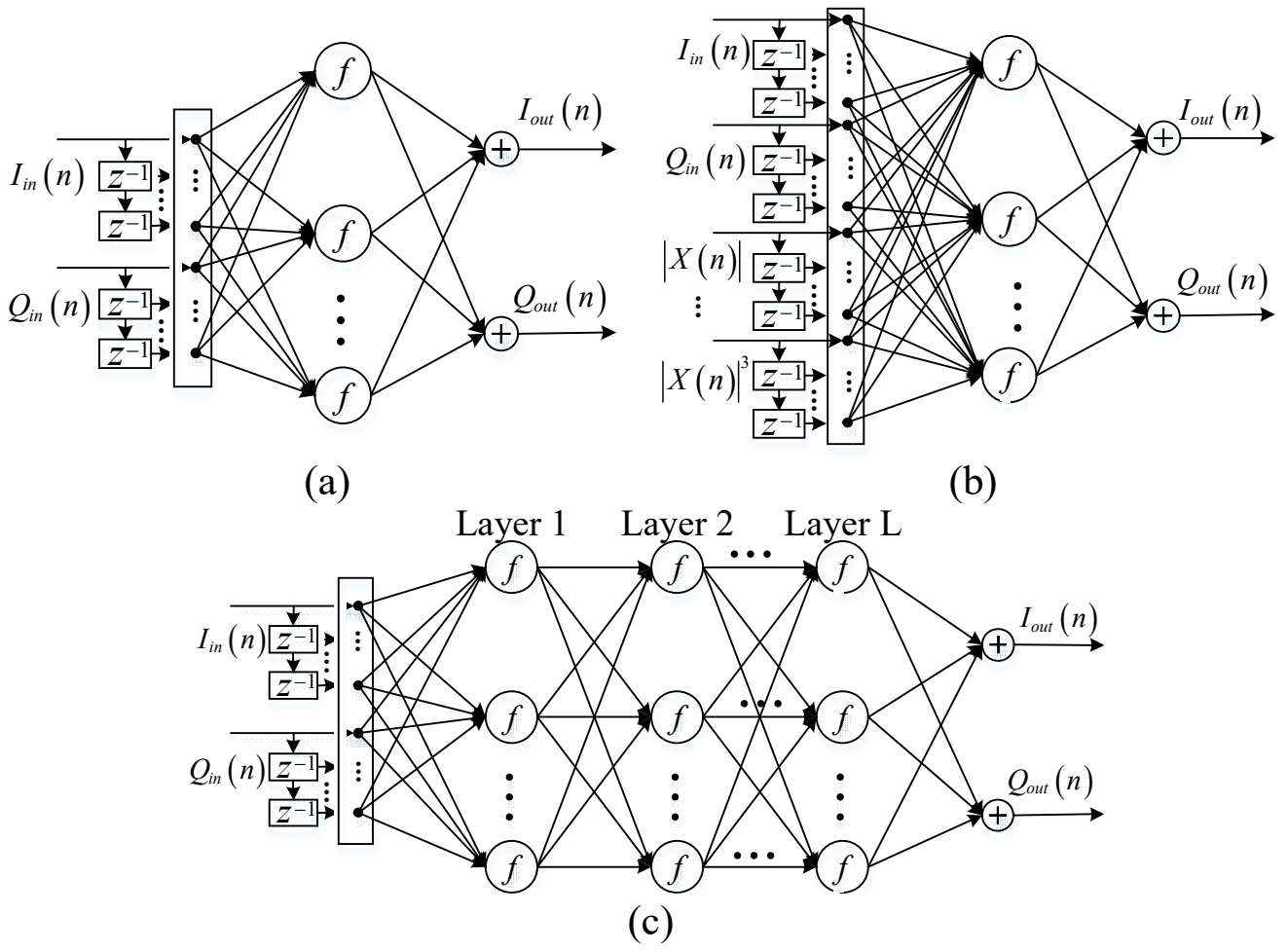

Fig. 1. Conventional ANN topologies for PA modeling. (a) Shallow NN Topology with input of the I/Q components. (b) Shallow NN Topology with input of the I/Q components and the envelope-dependent terms. (c) DNN Topology.

A commonly used shallow NN structure includes an input layer, a hidden layer structure with one or two layers, and an output layer. The model in Fig. 1 (a) considers injecting the in-phase and quadrature (I/Q) components of the input signal and embeds their corresponding time-delayed values into the spatial structure of the input layer of the network to reflect the corresponding memory effects, such as real-valued time-delay NN (RVTDNN) model in [15]. However, hidden and related information, such as envelope-dependent terms, requires further network computation capability, which leads to a complex network structure and additional hidden layers. To this end, the structure in Fig. 1(b) is proposed to simplify the network structure, which attempts to inject I/Q components and important envelope-dependent terms. The corresponding models include augmented radial basis function NN (ARBFNN) in [18] and augmented real-valued time-delay NN (ARVTDNN) in [19]. However, with the increase of signal bandwidth, the memory depth considered will also increase, and the input dimension of the model will grow significantly, resulting in a complex network structure. Overall, to provide sufficient network capacity, the shallow NN structure makes the calculation relatively complex.

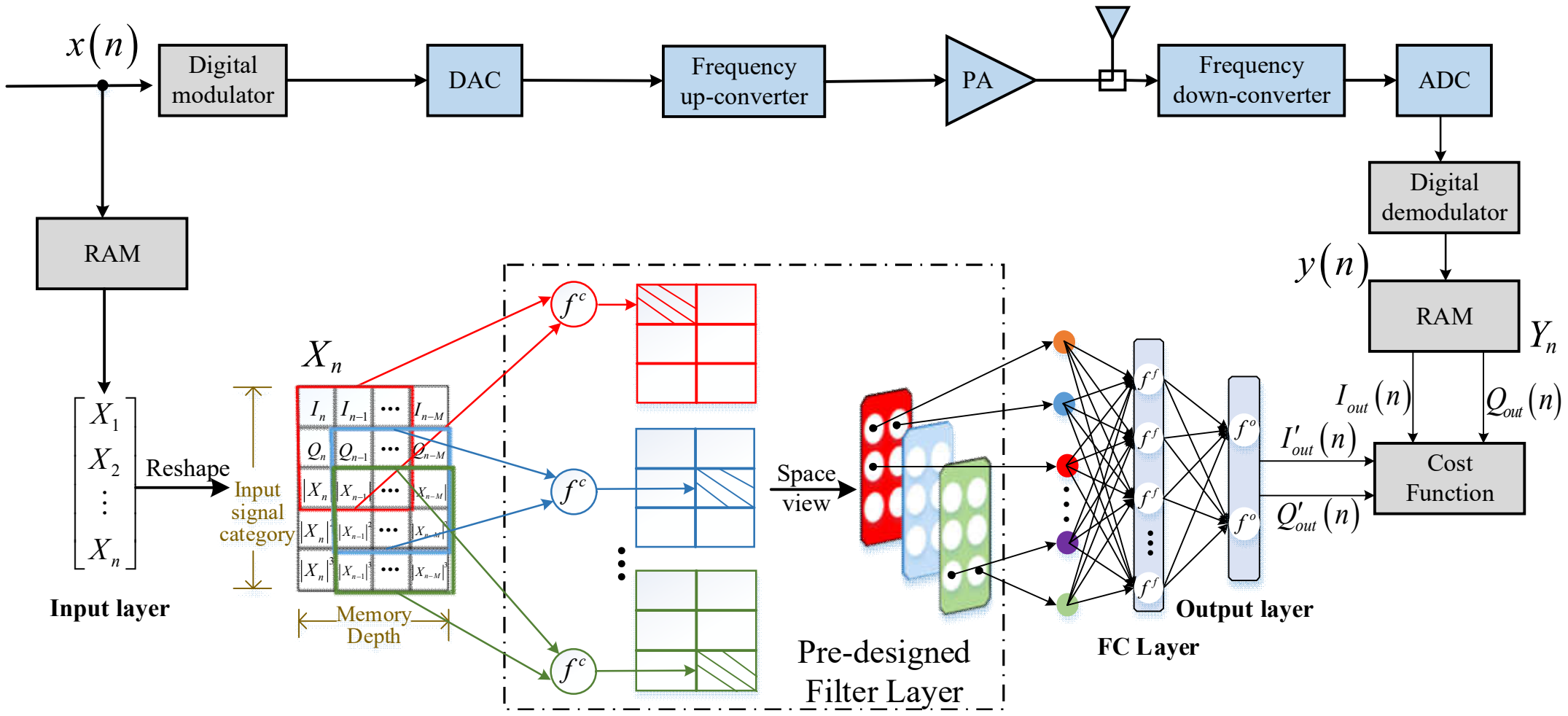

Fig. 2. Block diagram of proposed RVTDCNN model.

## *B. Deep Neural Networks for PA modeling*

NN with multiple hidden layers is proposed to improve the performance of PA modeling. Instead of the simple hidden layer, DNNs' architecture includes over three hidden layers to mimic and approximate the nonlinearity and memory effects of PA, as shown in Fig. 1(c). The corresponding models include the DNN model in [14]. As the number of the hidden layer increases, the fitting and generalization capabilities of the NN model increase [14], so it is fair to assume that the accuracy of modeling will increase with the number of the hidden layers. Different from the shallow neural networks, DNN can build more complex models with relatively low complexity. From the experiment conducted in [14], the networks can achieve the same accuracy with relatively low complexity. However, when PA exhibits complicated nonlinear characteristics and deep memory effects, it is difficult to achieve low complexity modeling performance with DNN. In addition, their implementation demands excessive signal processing resources as the signal bandwidth gets wider.

To further reduce the complexity of DNN, CNN and RNN are alternative methods. However, regression algorithms based on RNNs learning were originally designed for natural speech processing. If they are used for modeling and linearizing of the PAs, they seem to have high complexity. Also, the weight sharing structure of the CNN network has a remarkable effect in reducing the complexity of the model.

# III. REAL-VALUED TIME-DELAY CONVOLUTIONAL NEURAL NETWORK

The proposed RVTDCNN model structure is shown in Fig. 2. The RVTDCNN structure includes four layers, namely one input layer, one pre-designed filter layer, one fully connected (FC) layer, and one output layer. The pre-designed filter layer is constructed using a convolutional layer and is used to capture in an effective manner the important features and characteristics of the input data. Due to the characteristics of the weight sharing and data dimensionality reduction of the convolution kernel in the pre-designed filter structure, the input information can be extracted at a small network scale. The dimensions of each convolution kernel can be designed to yield low computation complexity while maintaining a good model's prediction performance. After the pre-designed filter layer, a fully connected layer is used to integrate valid features. The final output layer consists of two neurons with a linear activation function, corresponding to the I/Q components of the samples.

To construct the input graph of the convolutional network, the input data is a two-dimensional graph, including the I/Q components and the envelope-dependent terms of current and past signals. The input matrix is expressed as follows.

$$\begin{aligned} X_n = \big[ & I_{in}(n), I_{in}(n-1), \ldots, I_{in}(n-M); \\ & Q_{in}(n), Q_{in}(n-1), \ldots, Q_{in}(n-M); \\ & |x(n)|, |x(n-1)|, \ldots, |x(n-M)|; \\ & |x(n)|^2, |x(n-1)|^2, \ldots, |x(n-M)|^2; \\ & |x(n)|^3, |x(n-1)|^3, \ldots, |x(n-M)|^3 \big] \end{aligned} \quad (1)$$

where $I_{in}(n)$ and $Q_{in}(n)$ represent the I/Q components of the PA input signal $x(n)$, respectively; $|x(n)|$ denotes the amplitude of the current signal; $I_{in}(n-i)$, $Q_{in}(n-i)$ and $|x(n-i)|$, $(i=1,2,\cdots,M)$ denote the corresponding terms of past samples, respectively; $M$ represents the memory depth.

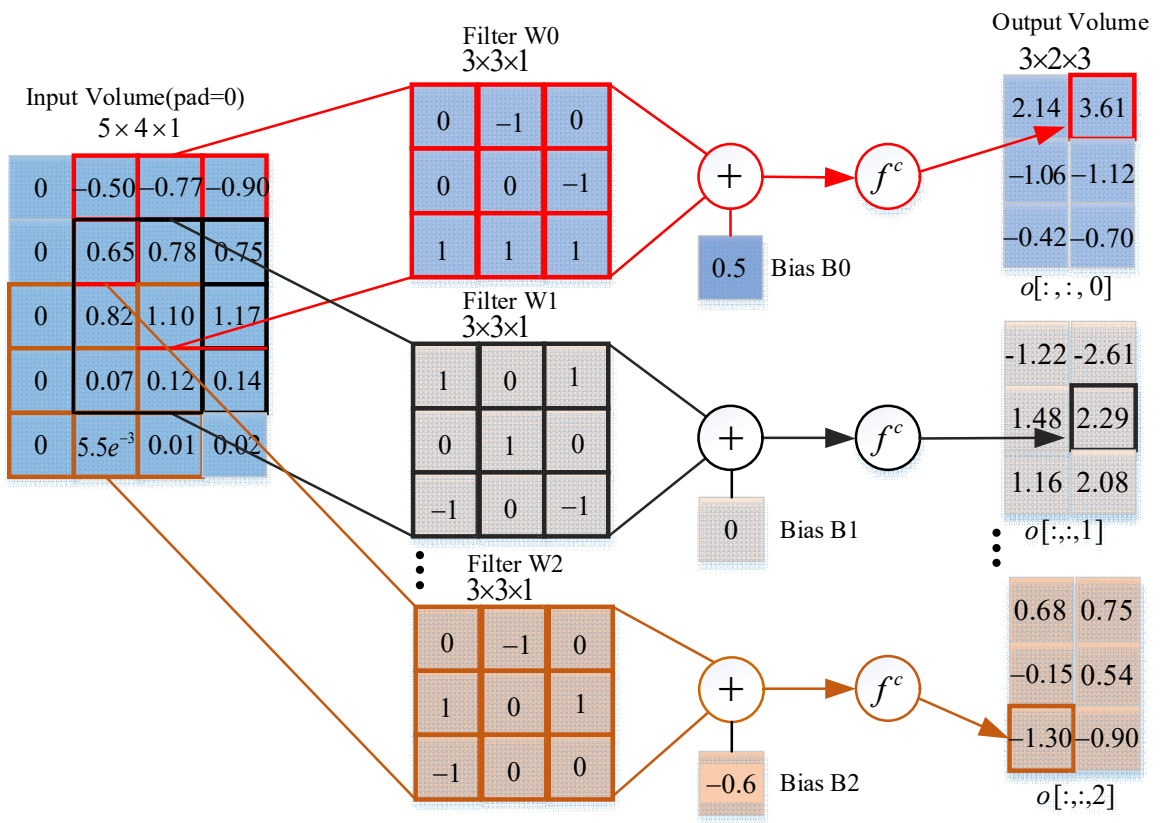

Fig.3 Two-dimensional convolution diagram.

The reason why the input data is arranged from a one-dimensional vector to a two-dimensional graph is to put it in a format suitable to the convolutional processing. The input data items corresponding to the adjacent delay signals are arranged adjacently, which ensures that the two-dimensional convolution kernel extracts the cross-terms of the differently delayed signals. As shown in Fig. 3, the input graph $X_n$ is transformed into a volume of the feature's map by pre-designed filter layer. This is accomplished by convolving the input data with the multiple local convolution kernels and adding bias's parameters to generate the corresponding local features, as shown in Fig. 4. The convolution operation is expressed as follows.

$$h_l = X_n \otimes \omega_l^c = \begin{bmatrix} I_{in}(n) & I_{in}(n-1) & \cdots & I_{in}(n-M) \\ Q_{in}(n) & Q_{in}(n-1) & \cdots & Q_{in}(n-M) \\ |x(n)| & |x(n-1)| & \cdots & |x(n-M)| \\ |x(n)|^2 & |x(n-1)|^2 & \cdots & |x(n-M)|^2 \\ |x(n)|^3 & |x(n-1)|^3 & \cdots & |x(n-M)|^3 \end{bmatrix} \otimes \omega_l^c \quad (2)$$

where $h_l,(l=1,2,...,L)$ represents the convolution output of the $l$-th convolution kernel with the input volume data $X_n$ arranged in 2D graph as illustrated in Fig. 3; $L$ represents the number of convolution kernels; $\omega_l^c$ represents the coefficients of the $l$-th convolution kernel; $\otimes$ shows the operation of convolution.

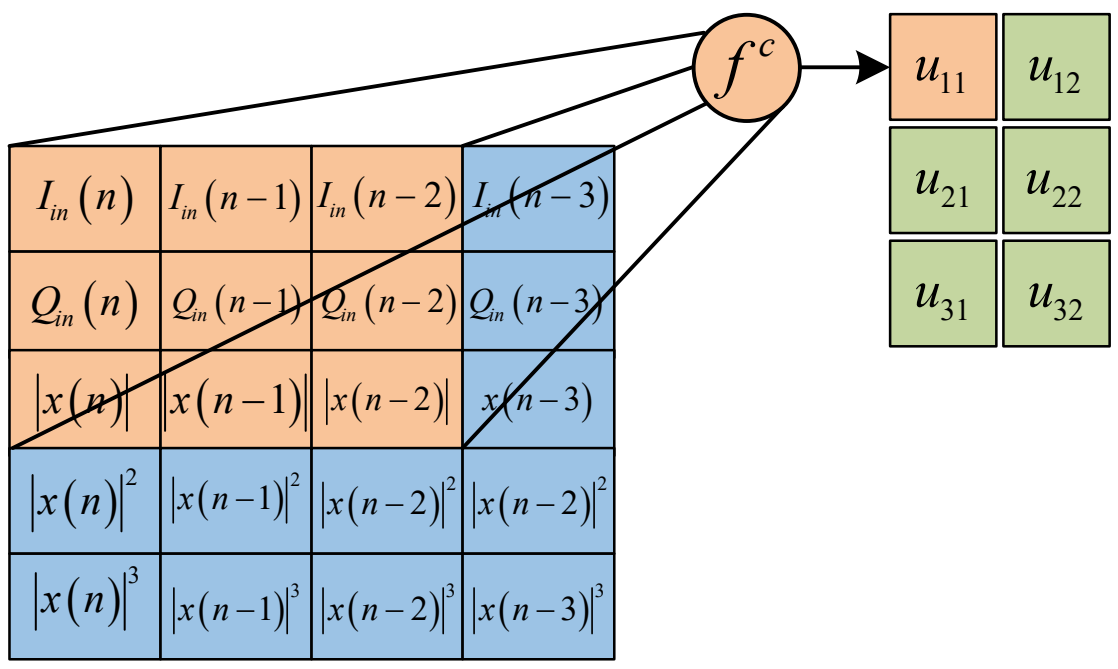

Fig. 4 The convolutional kernel extraction range example.

The output of the convolution process is then passed through a nonlinear activation function to obtain the nonlinear fitting ability. The outputs of the activation function of the convolution kernel are written as

$$u_l = f^c\left(h_l + b_l^c\right) \quad (3)$$

where $u_l,(l=1,2,...,L)$ is the output feature maps of the $l$-th convolution kernel; $f^c(\cdot)$ is the activation function of the convolution kernels. $b_l^c$ represents the bias of the $l$-th convolution kernels. Through the pre-designed filter, the rich basis function features required for PA modeling are extracted, which is proved in the appendix.

Then, the basis function features extracted by the pre-designed filter are arranged into a feature's vector to be injected into the FC layer. The feature's vector is expressed as

$$\begin{aligned} m &= \left[m_1, m_2, ..., m_{L\times B\times C}\right] \\ &= \left[u_1(1,1), u_1(1,2), ..., u_1(B,C), u_2(1,1), ..., u_L(B,C)\right] \end{aligned} \quad (4)$$

where $m$ is a vector of $L\times B\times C$; the dimension of feature's maps is $B\times C$.

The output of the FC layer is obtained as follows.

$$a_t = f^f\left(\sum_{i=1}^{L\times B\times C} \omega_{ti}^f m_i + b_t^f\right) \quad (5)$$

where $a_t,(t=1,2,...,T)$ denotes the $t$-th neuron output; $T$ represents the number of neurons in the FC layer; $\omega_{ti}^f$ and $b_t^f$ represent weights and bias, respectively; $f^f(\cdot)$ is the activation function in the FC layer.

Finally, the output layer weights and sums the output characteristics of the FC layer to acquire the network output. To ensure continuous values for the output data, we adjust the activation function, $f^o$, of the output layer by setting it as a linear function $y=x$.

$$\begin{cases} I'_{out}(n) = \sum_{t=1}^{T} \omega_{1t}^o a_t + b_1^o \\ Q'_{out}(n) = \sum_{t=1}^{T} \omega_{2t}^o a_t + b_2^o \end{cases} \quad (6)$$

where $I'_{out}(n)$ and $Q'_{out}(n)$ represent the neuron output in the output layer, which correspond to the prediction of the I/Q components of the output sample. $\{\omega_{1t}^o, \omega_{2t}^o, b_1^o, b_2^o\}$ represents the weights and biases of the output layer.

The label data contains I/Q components of the output samples. The output data vector is represented as

$$Y_n = \left[I_{out}(n), Q_{out}(n)\right]^T \quad (7)$$

where $I_{out}(n)$ and $Q_{out}(n)$ represent the I/Q components of the PA output signal $y(n)$, respectively.

## IV. Analysis of RVTDCNN Behavior Model

### *A. Training, Validation, and Testing of RVTDCNN*

The input and output signals of PA are sampled and saved in the random access memory (RAM), and then the RVTDCNN model is trained. We first train all parameters $\theta_k = \{\omega_k^c, b_k^c, \omega_k^f, b_k^f, \omega_k^o, b_k^o\}$ of the RVTDCNN model with the Adam optimization algorithm [44]. The trained convolutional layer is used as a pre-designed filter to extract the features of the input data. Due to the extraction effect of the pre-designed filter on the basis function, only a simple fully connected layer can be used to fit the behavioral characteristics of the PA. Therefore, during the modeling, to fix the parameters of the pre-designed filter, only the parameters $\theta_k^f = \{\omega_k^f, b_k^f, \omega_k^o, b_k^o\}$ of the fully connected layer and the output layer need to be adjusted to use the Levenberg- Marquardt (LM) algorithm [45].

TABLE II
THE MODELING PERFORMANCES UNDER DIFFERENT CONVOLUTION KERNEL SCALES

| Series No. | Size of conv kernel | Num. of conv kernel | Num. of model coef | NMSE (dB) | ACPR (dB) (-/+20MHz) | Series No. | Size of conv kernel | Num. of conv kernel | Num. of model coef | NMSE (dB) | ACPR (dB) (-/+20MHz) |
|---|---|---|---|---|---|---|---|---|---|---|---|
| 1 | 2*1*1 | 1 | 385 | -34.29 | -44.49/-44.51 | 13 | 3*1*1 | 1 | 306 | -33.81 | -44.21/-43.79 |
| 2 | 2*1*1 | 2 | 708 | -35.90 | -45.12/-45.33 | 14 | 3*1*1 | 2 | 550 | -34.98 | -45.11/-44.76 |
| 3 | 2*1*1 | 3 | 1031 | -36.63 | -45.89/-46.04 | 15 | 3*1*1 | 3 | 794 | -36.43 | -45.68/-45.74 |
| 4 | 2*1*1 | 4 | 1354 | -36.72 | -45.86/-46.13 | 16 | 3*1*1 | 4 | 1038 | -36.30 | -45.32/-45.79 |
| 5 | 2*2*1 | 1 | 307 | -32.52 | -43.96/-43.83 | 17 | 3*2*1 | 1 | 249 | -32.19 | -43.44/-43.48 |
| 6 | 2*2*1 | 2 | 552 | -36.22 | -45.41/-45.59 | 18 | 3*2*1 | 2 | 436 | -36.27 | -45.38/-45.57 |
| 7 | 2*2*1 | 3 | 797 | -36.42 | -45.59/-45.96 | 19 | 3*2*1 | 3 | 623 | -36.48 | -45.68/-45.92 |
| 8 | 2*2*1 | 4 | 1042 | -36.39 | -45.57/-45.78 | 20 | 3*2*1 | 4 | 810 | -36.39 | -45.59/-45.67 |
| 9 | 2*3*1 | 1 | 229 | -32.34 | -43.68/-43.54 | 21 | 3*3*1 | 1 | 192 | -32.16 | -43.31/-43.58 |
| 10 | 2*3*1 | 2 | 396 | -36.41 | -45.68/-45.87 | 22 | 3*3*1 | 2 | 322 | -35.07 | -45.19/-44.88 |
| 11 | 2*3*1 | 3 | 563 | -36.29 | -45.44/-45.32 | 23 | 3*3*1 | 3 | 452 | -36.64 | -45.99/-45.94 |
| 12 | 2*3*1 | 4 | 730 | -36.42 | -45.55/-45.87 | 24 | 3*3*1 | 4 | 582 | -36.71 | -45.87/-46.21 |

The goal of network training is to minimize the error between the label (measured output) data and the RVTDCNN model output determined in the forward path by updating the parameters of epoch $k$ until convergence of the network. In the forward path, we define the mean square error (MSE) as a cost function, which can be expressed as

$$E_{mse}(\theta) = \frac{1}{2N}\sum_{n=1}^{N}\left[\left(I'_{out}(n) - I_{out}(n)\right)^2 + \left(Q'_{out}(n) - Q_{out}(n)\right)^2\right] \quad (8)$$

where $I'_{out}(n)$ and $Q'_{out}(n)$ represent the output of the RVTDCNN model, respectively; $I_{out}(n)$ and $Q_{out}(n)$ represent the I/Q components of the output samples, respectively; $N$ is the length of the training data.

In this paper, 7,000 sets of modeling data are used for the modeling of RVTDCNN. Each set of modeling data contains input data and label data (measured output). The input data is a two-dimensional graph with a dimension of $5 \times M$ , where $M$ is the memory depth, as shown in Eq. (1). The label data is a vector with a dimension of $2 \times 1$ and is composed of the I/Q components of the PA output, as shown in Eq. (7). We divide the modeling data into the training set and test set according to the ratio of 3:2. Therefore, the training set contains 4,200 sets of modeling data, and the test set contains 2,800 sets of modeling data. The training set is used to train the model, and the unseen test set is used to test the final model to verify the generalization ability of the model. The modeling performance is described by normalized mean square error (NMSE).

$$NMSE = 10 \times \lg \frac{\frac{1}{N}\sum_{n=1}^{N}\left(\left(I'_{out}(n) - I_{out}(n)\right)^2 + \left(Q'_{out}(n) - Q_{out}(n)\right)^2\right)}{\frac{1}{N}\sum_{n=1}^{N}\left(\left(I_{out}(n)\right)^2 + \left(Q_{out}(n)\right)^2\right)} \quad (9)$$

In the Adam optimization algorithm, the initialization parameters $\beta_1$ and $\beta_2$ are used to control the exponential decay rate of moving averages of the gradient and the squared gradient, which are required to be close to 1. Through experimental verification, this paper sets the parameters to default values $\beta_1 = 0.9, \beta_2 = 0.999$ . Initialized 1st moment vector $\mu_0$ and 2st moment vector $\upsilon_0$ are often set to $\mu_0 = 0$ , $\upsilon_0 = 0$ . The constant $\varepsilon$ is used to prevent 2st moment vector from being 0. This paper sets $\varepsilon$ to the default value of $10^{-8}$ . We analyzed the cost function values and corresponding NMSE performance at different learning rates, as shown in Table I. It can be found from Table I that when the learning rate is $1\times10^{-3}$ , the NMSE performance is almost optimal, and the corresponding MSE is $1.24\times10^{-7}$ . At this time, the learning rate is also the choice of the fastest training speed at the best performance. Therefore, the learning rate is set to $1\times10^{-3}$ , and the threshold for the cost function is set to $1.2\times10^{-7}$ . The training process of the RVTDCNN model is shown in algorithm 1.

TABLE I
THE COST FUNCTION AND NMSE AT DIFFERENT LEARNING RATES

| Learning Rate | $1\times10^{-2}$ | $5\times10^{-3}$ | $1\times10^{-3}$ | $5\times10^{-4}$ | $1\times10^{-4}$ |
|---|---|---|---|---|---|
| MSE | $1.75\times10^{-7}$ | $1.49\times10^{-7}$ | $1.24\times10^{-7}$ | $1.14\times10^{-7}$ | $1.03\times10^{-7}$ |
| NMSE (dB) | -35.13 | -35.82 | -36.44 | -36.55 | -36.63 |

**Algorithm 1** Training of the RVTDCNN model

**Definition:**

1. Determine the network structure of the RVTDCNN model;
2. Get 4,200 sets of training data including input data and label data;
3. Define the cost function $E_{mse}(\theta)$ of the RVTDCNN network;
4. Define the convergence threshold $E_0 = 1.2\times10^{-7}$ of the cost function.

**Extraction of the Pre-designed Filter:**

1. Initialization:
 1) Set the learning rate $\alpha = 10^{-3}$ , and the exponential decay rates $\beta_1 = 0.9, \beta_2 = 0.999$ ;
 2) Initialize 1st moment vector $\mu_0 = 0$ and 2st moment vector $\upsilon_0 = 0$ ;
 3) Set the constant $\varepsilon = 10^{-8}$ ;
2. Training the RVTDCNN model:
 **Loop**: k=1, 2, …, 20,0000
 1) Calculate the network output from Eq. (6) and the cost function from Eq. (8);
 2) Judgment: if performance requirements are met, exit the loop;
 3) Calculate partial derivative of the objective function to coefficients $g_k = \frac{\partial E_{mse}(\theta_{k-1})}{\partial \theta_{k-1}}$ ;
 4) Update biased first and second moment estimate $\mu_k = \beta_1 \cdot \mu_{k-1} + (1-\beta_1)\cdot g_k$ , $\upsilon_k = \beta_2 \cdot \upsilon_{k-1} + (1-\beta_2)\cdot g_k^2$ ;
 5) Get bias-corrected first and second moment estimate $\hat{\mu}_k = \mu_k / (1-\beta_1^k)$ , $\hat{\upsilon}_k = \upsilon_k / (1-\beta_2^k)$ ;

6) Update coefficients $\theta_k = \theta_{k-1} - \alpha \cdot \hat{\mu}_k / \left(\sqrt{\hat{\upsilon}_k} + \varepsilon\right)$.
3. Save convolutional layer coefficients $\theta_c$ and define them as pre-designed filter coefficients.

**PA Modeling:**

Training the RVTDCNN model:

**Loop**: l=1, 2, …, 200

1) Calculate the pre-designed filter output from Eq. (4) using the coefficients $\theta_c$ ;
2) Calculate model output and cost function;
3) Judgment: if performance requirements are met, exit the loop;
4) Update network coefficients using LM algorithm;

**End**

To get the desired modeling performance, we need to decide the specific parameters of RVTDCNN. The 100 MHz OFDM input signal is taken as an example for description. The peak to average power ratio (PAPR) of the OFDM signal is 10.4 dB. The test PA is a Doherty PA. The small-signal gain of the PA is 28 dB, and the saturation power is 44 dBm. The choice of input data affects modeling performance and model complexity. An inappropriate input dimension of input data will increase the model coefficients. According to paper [19], the combination of the components $I$, $Q$, $|x(n)|, |x(n)|^2, |x(n)|^3$ is the best choice of the input signal to the NN yielding low model complexity and good performance. Based on the determined input data, the appropriate size of the convolution kernel becomes a factor affecting the modeling performance. The modeling performance and ACPR performance in DPD under different sizes and number of the convolution kernels was verified, and the results are shown in Table II. To decouple the effects of the FC layer and the pre-designed filter settings, the number of neurons in the FC layer is set to 20, which can provide sufficient network modeling capacity at different convolution kernel sizes. At this time, the modeling performance of the RVTDCNN model is only affected by the number and size of convolution kernels. It was found that the convolution kernels number significantly affects the modeling performance. If the convolution kernel number is equal to or less than 2, the model's NMSE performance increases with the increase of the convolution kernel number, regardless of the size of the convolution kernel. If the convolution kernel number exceeds 3, the NMSE performance does not increase significantly with the increase of the convolution kernel number. This can be explained by the fact that few convolution kernels cannot fully extract the features that reside in the input data. Meanwhile, when the convolution kernel number is kept constant, the size of the convolution kernel significantly affects the RVTDCNN model coefficient number. When the size of the convolution kernel is 3*3*1, the number of model coefficients is relatively small, and the NMSE performance is also quite good. The ACPR performance shows the same trend. Therefore, considering the modeling performance and model complexity, the convolution layer contains 3 convolution kernels of 3*3*1. The results in Table II correspond to the PA used in this paper. For different PAs, the optimal size and number of convolution kernels can be obtained through the scheme in our paper.

The neuron number in the FC layer is also an important factor affecting modeling performance and model complexity. To obtain the minimum number of neurons in the FC layer that can achieve the required performance, based on the determined input data and pre-designed filter structure, the modeling performance under the different neuron number in the FC layer was verified, and the results are shown in Fig. 5. It can be found that when the number of neurons is less than 6, the NMSE performance of the model will drop dramatically, meaning that few neurons cannot provide the required network modeling capacity. When the number of neurons is greater than 6, the NMSE performance of the model will not be significantly improved. Considering the model complexity and modeling performance, the neuron number in the FC layer was determined to be 6.

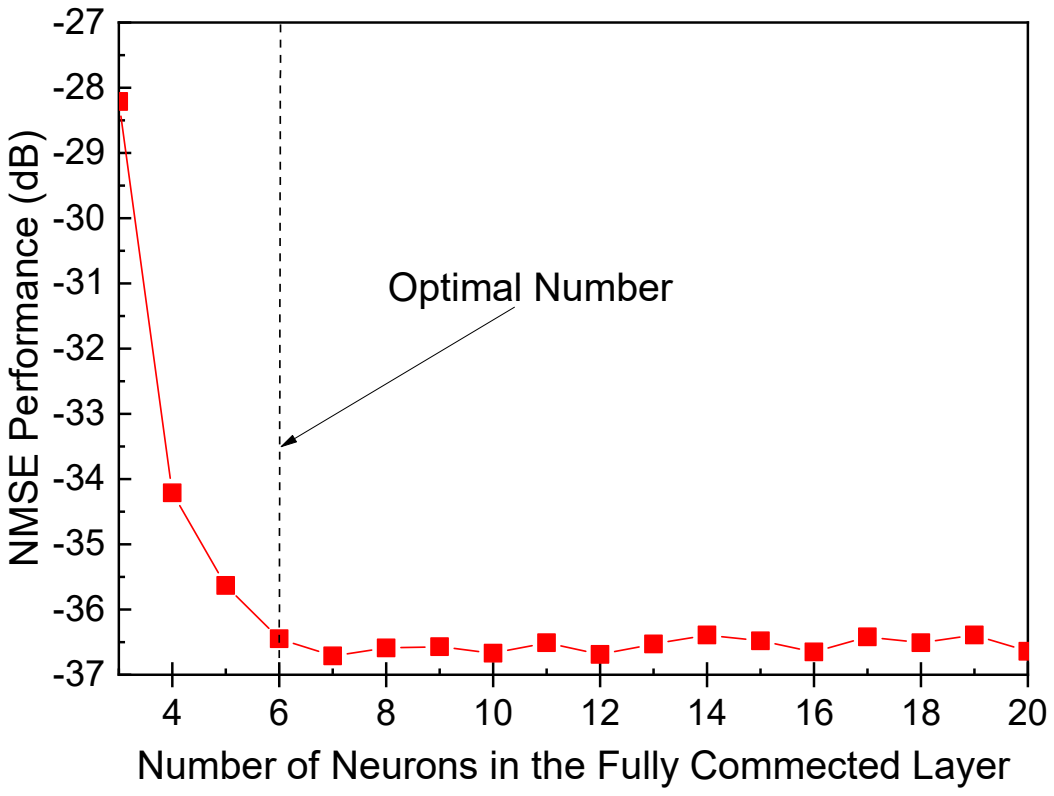

Fig. 5 NMSE performance under different neuron number in FC layer.

The activation functions commonly used in CNN are the sigmoid function, the Rectified Linear Unit (ReLU) function, the exponential linear unit (elu), the Leaky ReLU and the hyperbolic tangent sigmoid (Tanh) function, which are respectively defined in Eq. (10). To get the best modeling performance, the functions mentioned above were used to train the RVTDCNN model, and the results are shown in Table III, which can be summarized as the Tanh function solves the problem better than others.

$$
\begin{aligned}
&Sigmoid(x) = \frac{1}{1+e^{-x}} \\
&\mathrm{Re}LU(x) = \max(0, x) \\
&Elu(x) = \begin{cases} x, & if\ x \geq 0 \\ \alpha(e^x - 1), & if\ x < 0 \end{cases} \\
&Leaky_\mathrm{ReLU}(x) = \max(\lambda x, x) \\
&\tanh(x) = \frac{\exp(2x) - 1}{\exp(2x) + 1}
\end{aligned}
\tag{10}
$$

TABLE III
THE NMSE PERFORMANCE OF DIFFERENT ACTIVATION FUNCTION

| Activation function | NMSE(dB) |
|---|---|
| *sigmoid* | -34.75 |
| *ReLU* | -33.86 |
| *elu* | -35.21 |
| *Leakly ReLU* | -33.53 |
| *Tanh* | -36.44 |

## B. *Complexity analysis of RVTDCNN*

The complexity analysis aims to evaluate the capability of different models to assess if the training procedure of RVTDCNN is simpler than other typical model models. In terms of the model complexity, it refers to both the number of coefficients and the argument floating-point operations (FLOPs). The comparison of complexity, including the total number of coefficients of the network structure and the FLOPs, is shown in Table IV. Based on the theory and experimental data, RVTDCNN proposed in this paper has superior performance than traditional models for its convolution calculation. The following is a specific calculation process for the complexity of RVTDCNN.

TABLE IV
THE COMPLEXITY CALCULATIONS FOR DIFFERENT MODELS

| Model category | Model | The number of coefficients | | FLOPs(Computation complexity) | |
|---|---|---|---|---|---|
| | | Formula | Parameter Description | Formula | Parameter Description |
| Traditional model | GMP in [46] | $2K_aL_a + 2K_bL_bM_b + 2K_cL_cM_c$ | $K_a, L_a$ : index for aligned signal and envelope<br>$K_b, L_b, M_b$ : index for signal and lagging envelope<br>$K_c, L_c, M_c$ : index for signal and leading envelope | $8K_aL_a + 8K_bL_bM_b + 8K_cL_cM_c - 2$ | $K_a, L_a$ : index for aligned signal and envelope<br>$K_b, L_b, M_b$ : index for signal and lagging envelope<br>$K_c, L_c, M_c$ : index for signal and leading envelope |
| Shallow NNs | ARVTDNN in [19]<br>RVTDNN in [15] | $(N_i+1)N_1 + (N_1+1)N_o$ | $N_i$ : Number of neurons in the input layer<br>$N_1$ : Number of neurons in the hidden layer<br>$N_o$ : Number of neurons in the output layer<br>The number of hidden layers is 1 | $2N_iN_1 + 13N_1 + 2N_1N_o$ | $N_i$ : Number of neurons in the input layer<br>$N_1$ : Number of neurons in the hidden layer<br>$N_o$ : Number of neurons in the output layer<br>The number of hidden layers is 1 |
| DNN | DNN in [14] | $(N_i+1)N_1$<br>$\sum_{f=2}^{F}\left(N_{f-1}+1\right)N_f$<br>$+(N_F+1)N_o$ | $N_i$ : Number of neurons in the input layer<br>$N_f$ : Number of neurons in the $f$-th hidden layer<br>$F$ : The number of hidden layers<br>$N_o$ : Number of neurons in the output layer | $2N_iN_1 + 2N_FN_o$<br>$+2\sum_{f=2}^{F} N_{f-1}N_f$<br>$+15\sum_{f=1}^{F} N_f$ | $N_i$ : Number of neurons in the input layer<br>$N_f$ : Number of neurons in the $f$-th hidden layer<br>$F$ : The number of hidden layers<br>$N_o$ : Number of neurons in the output layer |
| RNN | LSTM in [41] | $4I(I+3)$<br>$+(I+1)N_1+$<br>$\sum_{f=2}^{F}\left(N_{f-1}+1\right)N_f$<br>$+(N_F+1)N_o$ | $I$ : Number of neurons in the LSTM layer<br>$N_f$ : Number of neurons in the $f$-th hidden layer<br>$F$ : The number of hidden layers<br>$N_o$ : Number of neurons in the output layer | $8IN_i +$<br>$I(M+1)(8I+71)$<br>$+2IN_1 + 2N_FN_O$<br>$+2\sum_{f=2}^{F} N_{f-1}N_f$ | $N_i$ : Number of neurons in the input layer<br>$M$ : Memory depth<br>$I$ : Number of neurons in the LSTM layer<br>$N_f$ : Number of neurons in the $f$-th hidden layer<br>$F$ : The number of hidden layers<br>$N_o$ : Number of neurons in the output layer |
| CNN | RVTDCNN | $P_{conv} + P_{fc} + P_{out}$ | $P_{conv}$ : Coefficient number of pre-designed filter<br>$P_{fc}$ : Coefficient number of fully connected layer<br>$P_{out}$ : Coefficient number of output layer | $FLOPs_{conv}$<br>$+FLOPs_{fc}$<br>$+FLOPs_{out}$ | $FLOPs_{conv}$ : FLOPs number of pre-designed filter<br>$FLOPs_{fc}$ : FLOPs number of fully connected layer<br>$FLOPs_{out}$ : FLOPs number of output layer |

Based on RVTDCNN, it can be stated that the convolution structure decreases the model size, which makes the extraction of the model's features more efficient. Compared to the standard feedforward network, the coefficient number of the convolutional structure to generate the same feature points is much less due to the weight sharing feature, which reduces the coefficient complexity of the RVTDCNN model. The total number of coefficients is equal to the sum of the weight number and the bias number between layers. Thus, the number of coefficients of the pre-designed filter layer can be calculated as follows.

$$\begin{aligned} P_{conv} &= W_{conv} + B_{conv} \\ &= r \times s \times z \times L + L \end{aligned} \tag{11}$$

where the kernel size is $r \times s \times z$, the number of kernels is $L$.

The coefficient number of the FC layer can be calculated as follows.

$$\begin{aligned} P_{fc} &= W_{fc} + B_{fc} \\ &= B \times C \times L \times T + T \end{aligned} \tag{12}$$

where $B \times C \times L$ denotes the size of the output tensor of the pre-designed filter, $T$ is the number of neurons of the FC layer.

The number of coefficients of the output layer can be obtained as follows.

$$\begin{aligned} P_{out} &= W_{out} + B_{out} \\ &= T \times T_{out} + T_{out} \end{aligned} \tag{13}$$

where $T_{out}$ represents the neuron number of the output layer.

In summary, the number of coefficients of RVTDCNN can be calculated as follows.

$$P_{TDCNN} = P_{conv} + P_{fc} + P_{out} \tag{14}$$

For a typical generalized memory polynomial (GMP) model, the complex coefficient number of the model represents the number of basis function terms considered. The number of real coefficients of the GMP model can be expressed as.

$$P_{GMP} = 2K_aL_a + 2K_bL_bM_b + 2K_cL_cM_c \tag{15}$$

where $K_a, L_a$ are the index for aligned signal and envelope; $K_b, L_b, M_b$ are the index for signal and lagging envelope; $K_c, L_c, M_c$ are the index for signal and leading envelope.

The ARVTDNN in [19], the RVTDNN in [15] and the DNN model in [14] are all fully connected networks. The coefficient number of the fully connected networks can be obtained as follows.

$$P_{MNNs} = \sum_{i_1=2}^{I_1}\left(N_{i_1-1}+1\right) \times N_{i_1} \tag{16}$$

where $N_{i_1}$ means the neurons of the $i_1$-th layer, $I_1$ means the number of layers ($I_1 \geq 3$, including the input and output layer).

For the LSTM model in [41], the number of model coefficients of the LSTM layer is

$$P_{LSTM} = 4I\left(N_{in} + I + 1\right) \tag{17}$$

where $N_{in}$ means the input number of the LSTM layer at each moment; $I$ is the number of neurons in the LSTM layer.

Except for the total number of the coefficients, the argument FLOPs are also introduced to assess the network complexity. For the convolutional process, considering the complexity of the activation function, the formula for calculating the number of FLOPs can be derived as follows.

$$FLOPs_{conv} = 2rsz \times BCL + 13BCL \tag{18}$$

For the FC layer in the RVTDCNN model, the FLOPs can be calculated as follows.

$$FLOPs_{fc} = 2\left(B \times C \times L \times T\right) + 13T \tag{19}$$

For the output layer in the RVTDCNN model, the FLOPs can be calculated as follows.

$$FLOPs_{out} = 4T \tag{20}$$

The FLOPs of the fully connected networks can be obtained as follows.

$$FLOPs_{MNNs} = \sum_{i_1=2}^{I_1} 2N_{i_1-1}N_{i_1} + K_{i_1}N_{i_1} \tag{21}$$

where $K_{i_1}$ is the FLOPs that calculate the activation function of the $i_1$-th layer.

For the LSTM model in [41], considering the complexity of the activation function, the FLOPs of the LSTM layer is

$$FLOP_{LSTM} = I\left(8N_{in} + 8I + 71\right) \tag{22}$$

According to the above formula, the calculation formulas of the complexity of the RVTDCNN model and other models are listed in Table VI.

## V. Experimental Setup

The experimental setup in Fig. 6 is used to evaluate the linearization performance of the proposed model. The test signal is a 100 MHz OFDM signal with a PAPR of 10.4 dB, which is generated by MATLAB on a personal computer (PC). The OFDM signal is compounded of multiple OFDM symbols, generated by 16-QAM symbols modulated onto 64 subcarriers and then filtered by a raised-cosine with the roll-off factor of 0.1. The test signal was first downloaded into the arbitrary waveform generator (AWG) 81180A. Then, the AWG transmits the generated baseband signal to the performance signal generator (PSG) E8267D through cable, which implements digital-to-analog (DAC) conversion and frequency-up conversion. The modulation frequency in PSG is 2.14-GHz. Then, the RF signal generated by PSG is fed into PA.

The PA output signal is fed into the coupler, whose output is connected to a high-power load. In the feedback loop, the output of the coupler is captured through the oscilloscope (MSO) 9404A. Then, the Keysight 89600 Vector Signal Analyzer (VSA) software running on the MSO analyzes the captured RF input signal, including frequency-down conversion and analog-to-digital (ADC) conversion. The sampling rate is set to 625-MHz. Then, the output baseband signal is captured by VSA and downloaded to the PC. The acquired input and output signals are processed in the Python software in a PC to construct the behavior model of PA. The 3.5.2 version of Python used is installed in the Windows environment, using the 2017.3.4 version of PyCharm as its integrated development environment.

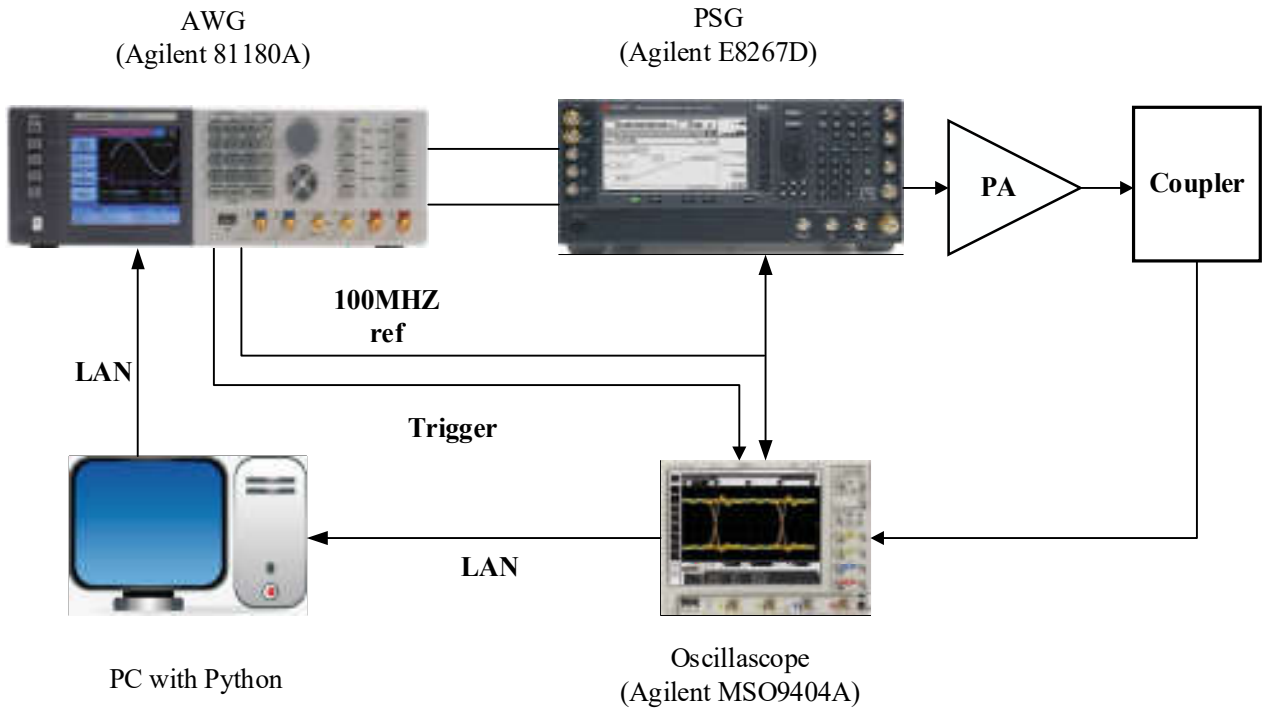

Fig. 6. Experimental setup.

## VI. Measurement Results

### A. Modeling Performance

RVTDCNN, as shown in Fig. 2, is used to illustrate the performance of this behavioral model. The proposed modeling method and other modeling methods are evaluated herein using the NMSE performance in Eq. (9) and ACPR performance in DPD. The optimal network structure of different methods for 100MHz is shown in Table V.

TABLE V
Network Structure of Different Methods ($M$=3).

| Parameter | Settings |
|---|---|
| **PA** | |
| Saturation power | 43 dBm |
| Frequency | 3.3-GHz~3.5-GHz |
| Center frequency | 3.4-GHz |
| Small-signal gain (SSG) | 28 dB |
| Output backoff (OBO) | 6 dB |
| Power added efficiency (PAE) | 50% |
| Output power at 1dB compression | 37 dBm |
| **RVTDCNN** | |
| Input data | $I/Q, \lvert X_n\rvert, \lvert X_n\rvert^2, \lvert X_n\rvert^3$ |
| Num. of input neuron | 5*4 |
| Num. of convolution kernels | 3 |
| Size of convolution kernels | 3*3*1 |
| Activation (Convolution layer) | 'Tanh' |
| Num. of neurons in FC layer | 6 |
| Activation (FC layer) | 'Tanh' |
| Num. of output neuron | 2 |
| **ARVTDNN** | |
| Input data | $I/Q, \lvert X_n\rvert, \lvert X_n\rvert^2, \lvert X_n\rvert^3$ |
| Num. of input neuron | 20 |
| Num. of neurons in the hidden layer | 17 |
| Activation (hidden layer) | 'Tansig' |
| Num. of output neuron | 2 |
| **RVTDNN** | |
| Input data | $I/Q$ |
| Num. of input neuron | 8 |
| Num. of neurons in the hidden layer | 35 |
| Activation (hidden layer) | 'Tanh' |
| Num. of output neuron | 2 |
| **DNN** | |
| Input data | $I/Q$ |
| Hidden layer structure | [17 17 17] |
| Activation (hidden layer) | 'Sigmoid' |
| Num. of output neuron | 2 |
| **GMP** | |
| The index arrays for aligned signal and envelope | $K_a$=11, $L_a$=7 |
| The index arrays for signal and lagging envelope | $K_b$=3, $L_b$=2, $M_b$=5 |
| index arrays for signal and leading envelope | $K_c$=2, $L_c$=0 $M_c$=3 |
| **LSTM** | |
| Input data | $I/Q$ |
| Num. of input neuron | 8 |
| Num. of neurons in the LSTM layer | 8 |
| Num. of neurons in FC layer | [7,5] |
| Activation (FC layer) | 'ReLU' |
| Num. of output neuron | 2 |

Fig. 7 shows the convergence curve of the RVTDCNN model training process. As shown in Algorithm 1, the threshold value of the cost function is set to $1.2\times10^{-7}$. When the cost function value of the network is less than the threshold, the network converges. It was found that the model convergence requires only 83 iterations (LM algorithm), so the RVTDCNN model has less training complexity. At the same time, the model converges synchronously on the training set and the test set, and MSE is almost the same, despite that the test data has never been used in training. Therefore, the model does not have overfitting problems. The NMSE performance of the model on the training set and test set is about -36.4 dB. Therefore, RVTDCNN has good generality for PA modeling.

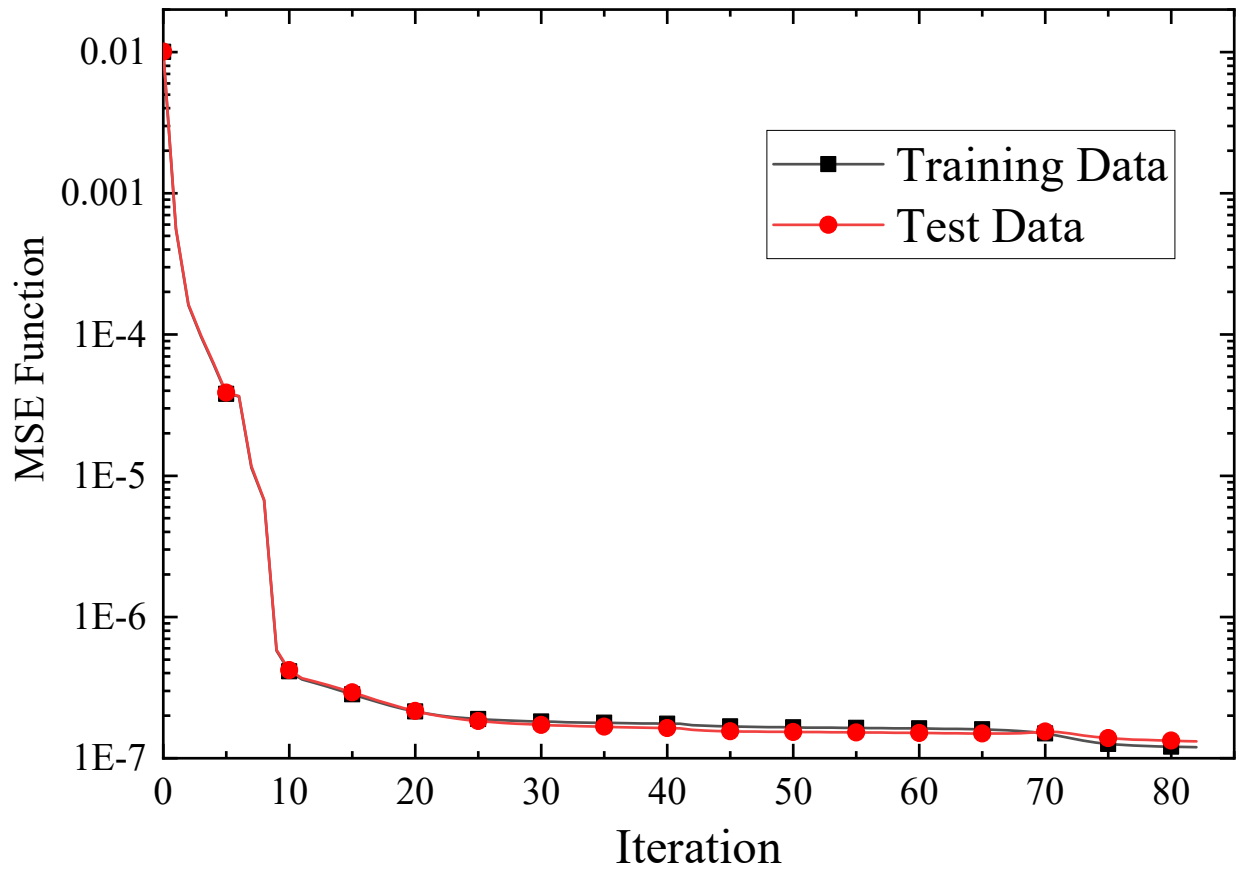

Fig. 7. Convergence curve of the training process.

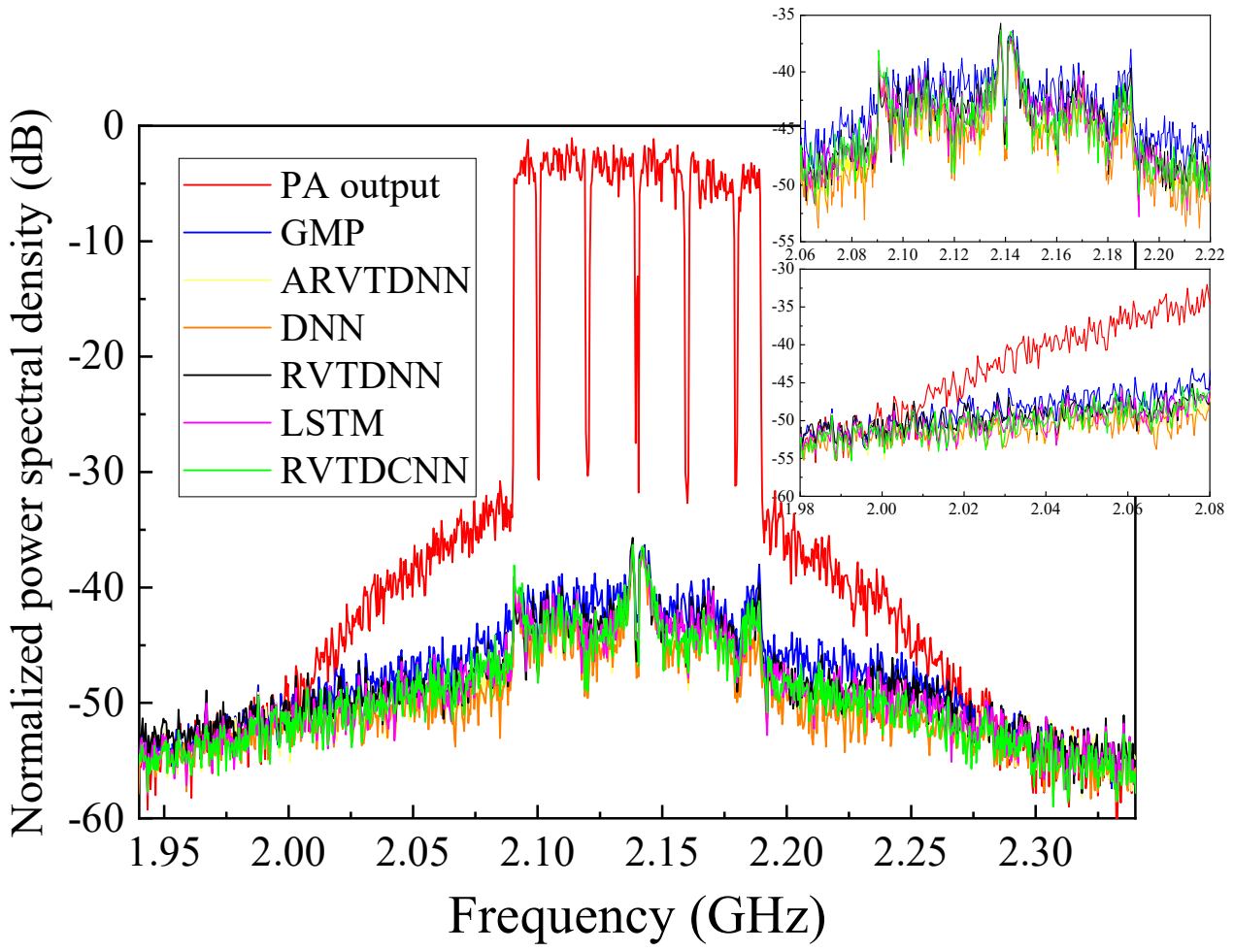

Fig. 8. Spectral comparison of modeling errors between RVTDCNN and other models at 100 MHz OFDM source signal.

Fig.8 compares the spectrum of modeling errors between the RVTDCNN model and other typical models at 100 MHz OFDM source signal. Modeling error is defined as the difference between model prediction output and PA output, which represents the modeling accuracy of the model. It can be seen from the figure that the error spectrum of RVTDCNN model is lower than -40 dB both out-of-band and in-band, which shows the effectiveness of the proposed method. Meanwhile, the error spectrum of the RVTDCNN model is lower than that of the GMP and RVTDNN model both out-of-band and in-band. Compared with the DNN model, the ARVTDNN and the LSTM model, the modeling performance of the proposed RVTDCNN model is also not reduced, which verifies the superiority of the proposed method in modeling performance.

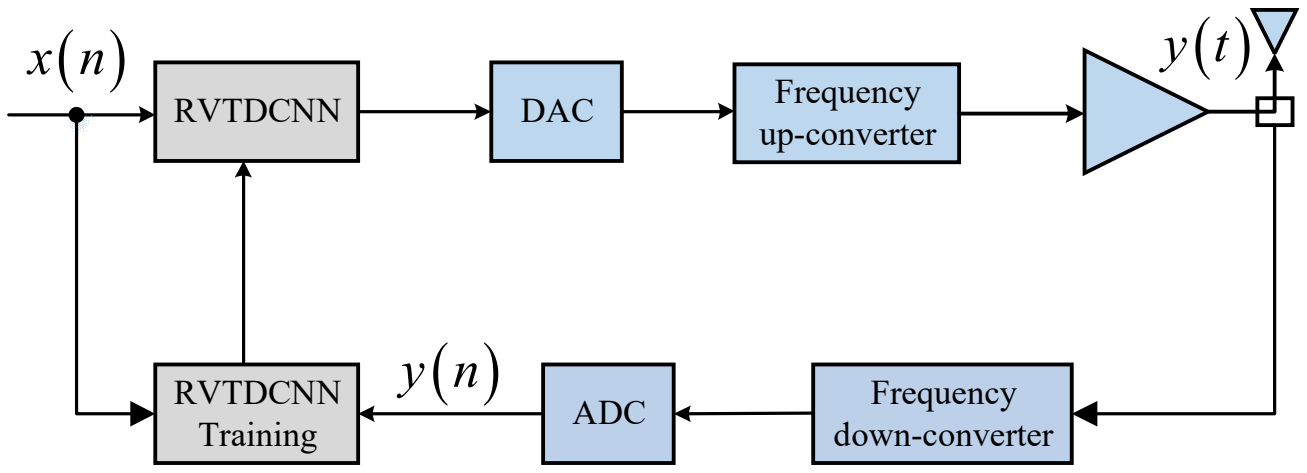

Fig. 9. The diagram of the proposed DPD architecture.

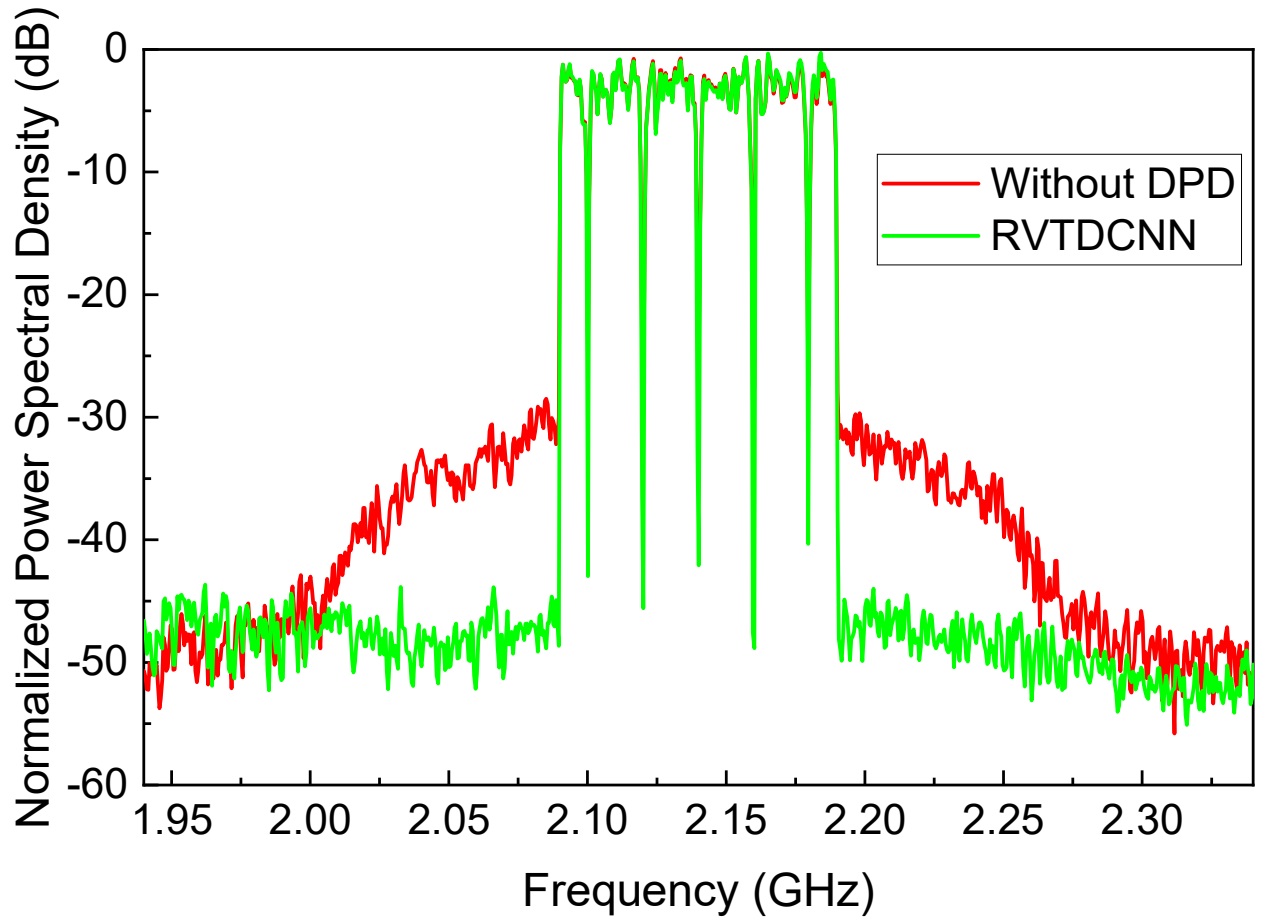

Fig. 10. Linearization performance of the PA using RVTDCNN at 100MHz OFDM source signal.

DPD is one of the most effective ways to alleviate the nonlinearity and memory effects of PA [2]. The DPD model is the inverse model of PA. Based on the indirect learning structure [19], the DPD can be implemented by placing the RVTDCNN model on the main path, as shown in Fig. 9. Then, the trained DPD model is used to update the DPD on the main path for the linearization of the PA. The input data of the DPD model is a two-dimensional matrix as shown in Eq. (1), including the I/Q components and envelope-dependent terms of the current and past signals of the PA output. The label data of the DPD model is the I/Q components of the input signal, as shown in Eq. (7). Take the signal of 100MHz as an example, when implementing DPD, the parameters and training methodology of the RVTDCNN model are the same as those for PA modeling.

Fig. 10 shows the output spectrum after the linearization of the PA using the RVTDCNN model at 100 MHz OFDM source signal. The same dimension was used to derive the inverse model, and it was found that the RVTDCNN inverse model (DPD model) has a significant effect in reducing the PA distortion when cascaded with the nonlinear PA. Using RVTDCNN to linearize the PA, the ACPR performance is reduced from -31 dBc to -46 dBc.

### *B. Comparison of Modeling Performance with Various Methods*

To prove the superiority of the RVTDCNN model, Table VI compares the NMSE performance and ACPR performance of DPD with the proposed modeling method and other methods, where the experimental results are based on the optimal dimension of the model structure. The structure and model dimensions of other models have been set to yield the best performance, as shown in Table V. The best NMSE value that the traditional GMP model can achieve is -33.19 dB, and the corresponding number of real model coefficients is 214. The results show that RVTDCNN can improve NMSE performance by about 3 dB compared to the traditional GMP model with about one-third less in the number of the model's coefficients when implemented in digital processors. Compared with the RVTDNN model, the RVTDCNN model can improve NMSE by about 1.3 dB with fewer model coefficients and FLOPs. It can be interpreted as that when the input contains only the I/Q components of the current and past signals, the single hidden layer network cannot generate sufficiently rich basis functions features, resulting in poor modeling performance [19]. Meanwhile, the RVTDCNN model can lead to almost the same NMSE performance with the ARVTDNN model, the DNN model and the LSTM model but with less than half of the required number of coefficients, as shown in Table VI.

TABLE VI
THE MODELING PERFORMANCE AND COMPLEXITY OF VARIOUS METHOD

| | FLOPs (/sample) | Num. of model coefficients | NMSE (dB) | ACPR (dB) (-/+20MHz) |
|---|---|---|---|---|
| GMP in [46] | 854 | 214 | -33.19 | -44.09/-43.65 |
| ARVTDNN in [19] | 1008 | 393 | -36.47 | -46.01/-45.29 |
| RVTDNN in [15] | 1155 | 387 | -35.09 | -45.74/-44.79 |
| DNN in [14] | 2306 | 801 | -36.42 | -45.63/-45.60 |
| LSTM in [41] | 5034 | 467 | -36.27 | -45.54/-45.49 |
| RVTDCNN | 876 | 158 | -36.44 | -45.87/-46.11 |

Fig.11 shows the complexity comparison between the RVTDCNN model and other models when obtaining similar modeling performance. At this time, the ARVTDNN model, DNN model, LSTM model, and the proposed RVTDCNN model have similar modeling performance, and the NMSE performance is about -36.30 dB. The GMP model and the RVTDNN model perform best modeling performance, and the NMSEs are -33.19 dB and -35.09 dB, respectively. It can be found that the proposed method has a significant decrease in the number of model coefficients and the number of FLOPs due to the weight sharing feature, while the modeling performance does not decrease.

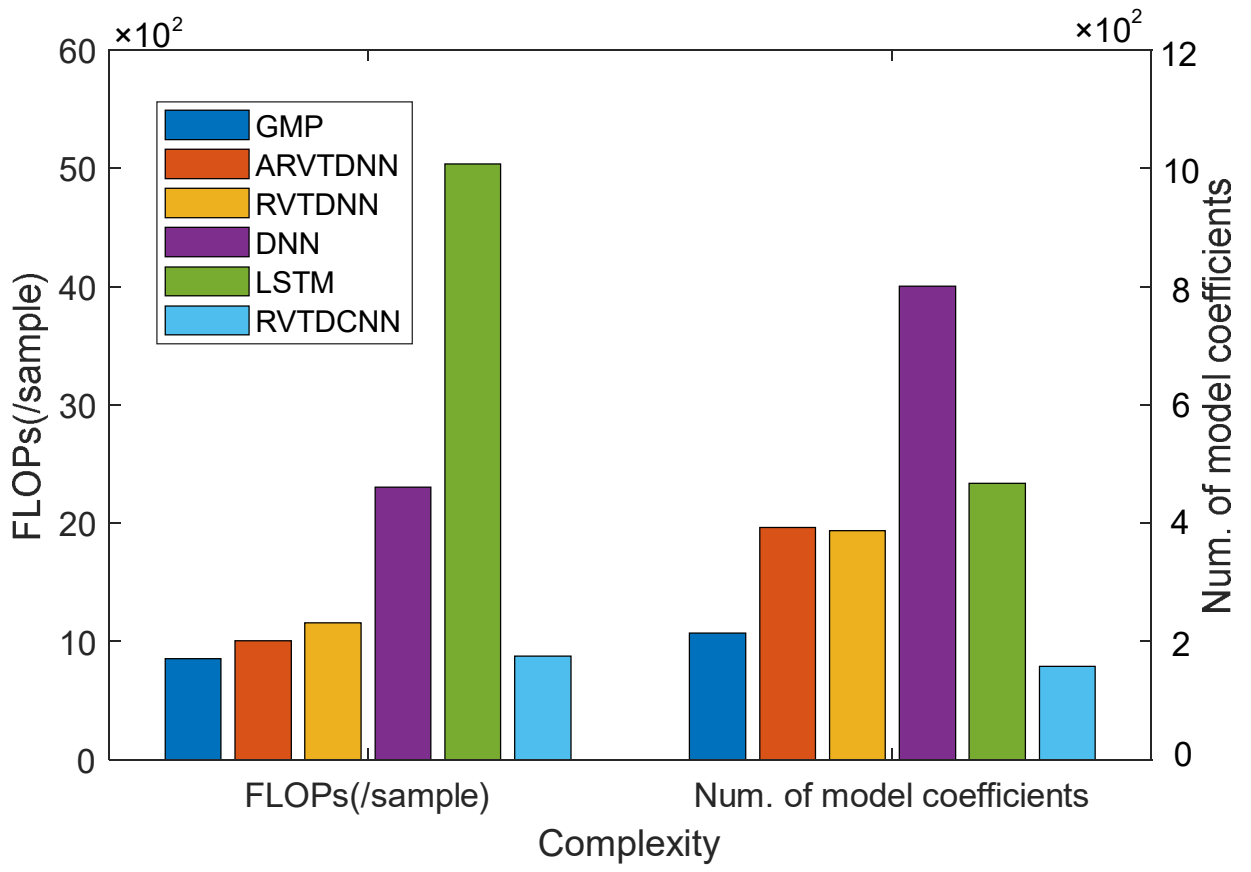

Fig. 11 Comparison of the complexity of the proposed RVTDCNN model and other models under similar modeling performance.

To further verify the performance of the RVTDCNN model under different transmitter's hardware impairment conditions, we evaluate three cases of the transmitter, as shown in Table VII. Case 1 indicates that only the nonlinear distortion of the PA is considered in the transmission chain. Case 2 indicates the existence of both the PA nonlinear distortion and I/Q imbalance. In Case 3, PA nonlinear distortion, I/Q imbalance and DC offset are included in the transmission chain. The specific distortion level is shown in Table VII.

TABLE VII
DIFFERENT CASED OF SIGNAL DISTORTION AT TRANSMITTER

| | Various Distortions | Distortion Level |
|---|---|---|
| Case 1 | PA nonlinearity: Yes<br>I/Q imbalance: No<br>DC offset: No | PA nonlinearity: 3 dB gain compression. |
| Case 2 | PA nonlinearity: Yes<br>I/Q imbalance: Yes<br>DC offset: No | PA nonlinearity: Same as Case 1;<br>I/Q imbalance:1 dB gain imbalance and 3-degree phase imbalance |
| Case3 | PA nonlinearity: Yes<br>I/Q imbalance: Yes<br>DC offset: Yes | PA nonlinearity: Same as Case 1;<br>I/Q imbalance: Same as Case 2;<br>DC offset: 3% for I and 5% for Q. |

Table VIII shows the NMSE and ACPR of the RVTDCNN model under different cases of the transmitter. It was found that RVTDCNN shows good modeling performance in case 1 because the system only contains the nonlinear distortion of the PA at this time. Meanwhile, RVTDCNN also shows superior performance in NMSE and ACPR performance for case 2 and case 3. The reason is that RVTDCNN can also eliminate the imperfections of transmitter besides PA's nonlinearity, such as DC offset and I/Q imbalance.

TABLE VIII
THE NMSE AND ACPR PERFORMANCE OF RVTDCNN MODEL UNDER DIFFERENT CASES

| | NMSE (dB) | ACPR (dB) (-/+20MHz) |
|---|---|---|
| Case 1 | -36.44 | -45.87/-46.11 |
| Case 2 | -36.07 | -45.57/-45.89 |
| Case 3 | -35.89 | -45.37/-45.48 |

Fig. 12 compares the gain and phase characteristics of the PA output between 100 MHz input signal and 200 MHz input signal. It can be found that the PA output of the 200 MHz input signal exhibits much stronger nonlinearity and memory effect than that of the 100M input signal.

TABLE IX
THE MODELING PERFORMANCE AND COMPLEXITY OF THE RVTDCNN MODEL UNDER DIFFERENT INPUT SIGNAL BANDWIDTH

| Input Signal Bandwidth | Memory Depth | ARVTDNN in [19] | | DNN model in [14] | | LSTM in [41] | | RVTDCNN model | |
|---|---|---|---|---|---|---|---|---|---|
| | | Num. of model coef | NMSE (dB) | Num. of model coef | NMSE (dB) | Num. of Model coef | NMSE (dB) | Num. of model coef | NMSE (dB) |
| 40-MHz | 2 | 308 | -36.78 | 767 | -36.60 | 467 | -36.66 | 104 | -36.87 |
| 60-MHz | 3 | 393 | -36.71 | 801 | -36.43 | 467 | -36.40 | 158 | -36.62 |
| 100-MHz | 3 | 393 | -36.47 | 801 | -36.42 | 467 | -36.27 | 158 | -36.44 |
| 200-MHz | 5 | 563 | -36.49 | 869 | -35.84 | 467 | -36.13 | 266 | -36.47 |

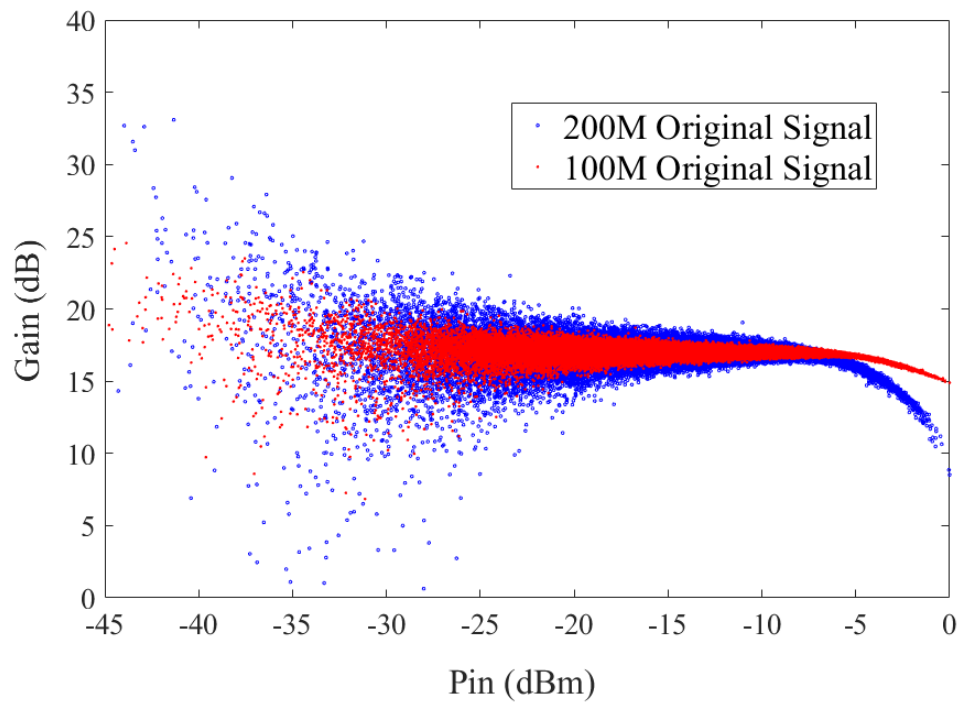

(a)

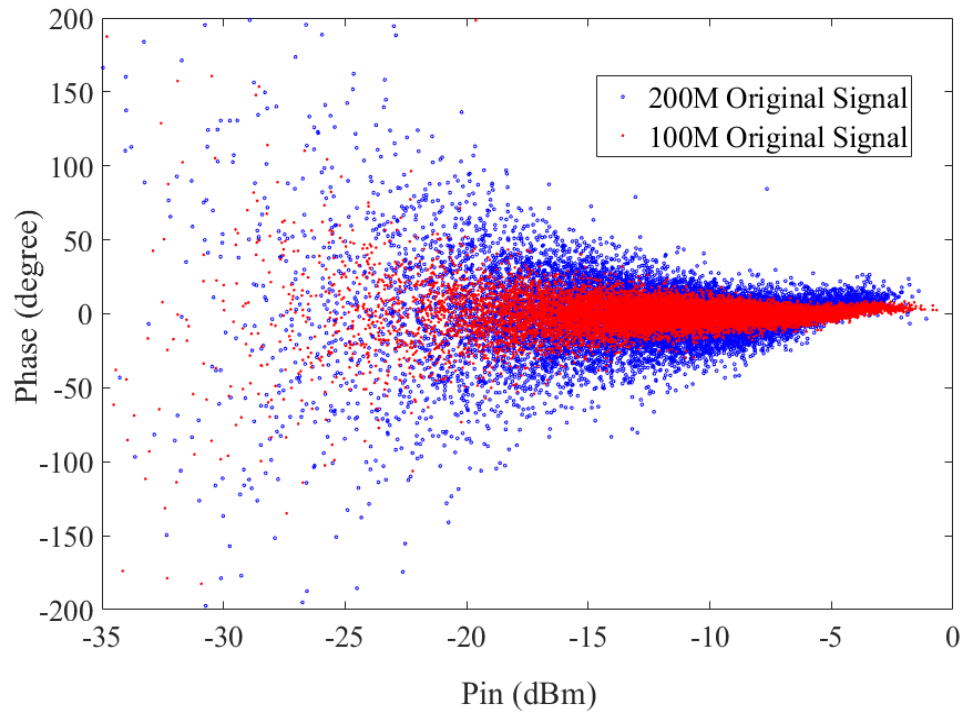

(b)

Fig. 12. Gain characteristics and phase characteristics of the PA output under different bandwidths. (a) Gain characteristics. (b) Phase characteristics.

Table IX shows the model complexity and modeling performance of RVTDCNN under different signal bandwidths. It can be deduced that as the signal bandwidth increases, the length of memory required for modeling increases accordingly. The same network structure is used to model PAs with different signal bandwidths, resulting in good modeling performance. If the signal bandwidth is further increased, we can achieve good modeling performance by increasing the number of convolution kernels and neurons in the fully connected layer. It can be found that for the traditional ANN models, such as ARVTDNN, the strong memory effect leads to a rapid increase in the model complexity. The memory depth increased from 2 to 5, and the number of the model coefficients of ARVTDNN increases to 563. For the proposed model, the memory depth increases from 2 to 5, and the number of model's coefficients is 266 that can be considered as reasonable, which is half the number of model coefficients of the ANN model. The coefficient number of the LSTM model does not increase with the signal bandwidth, but it is still about twice the coefficient number of RVTDCNN model. In other words, with the increase in signal bandwidth, the proposed model has more advantages in model complexity and fewer model coefficients. At the same time, the ACPR performance of RVTDCNN also verified this conclusion.

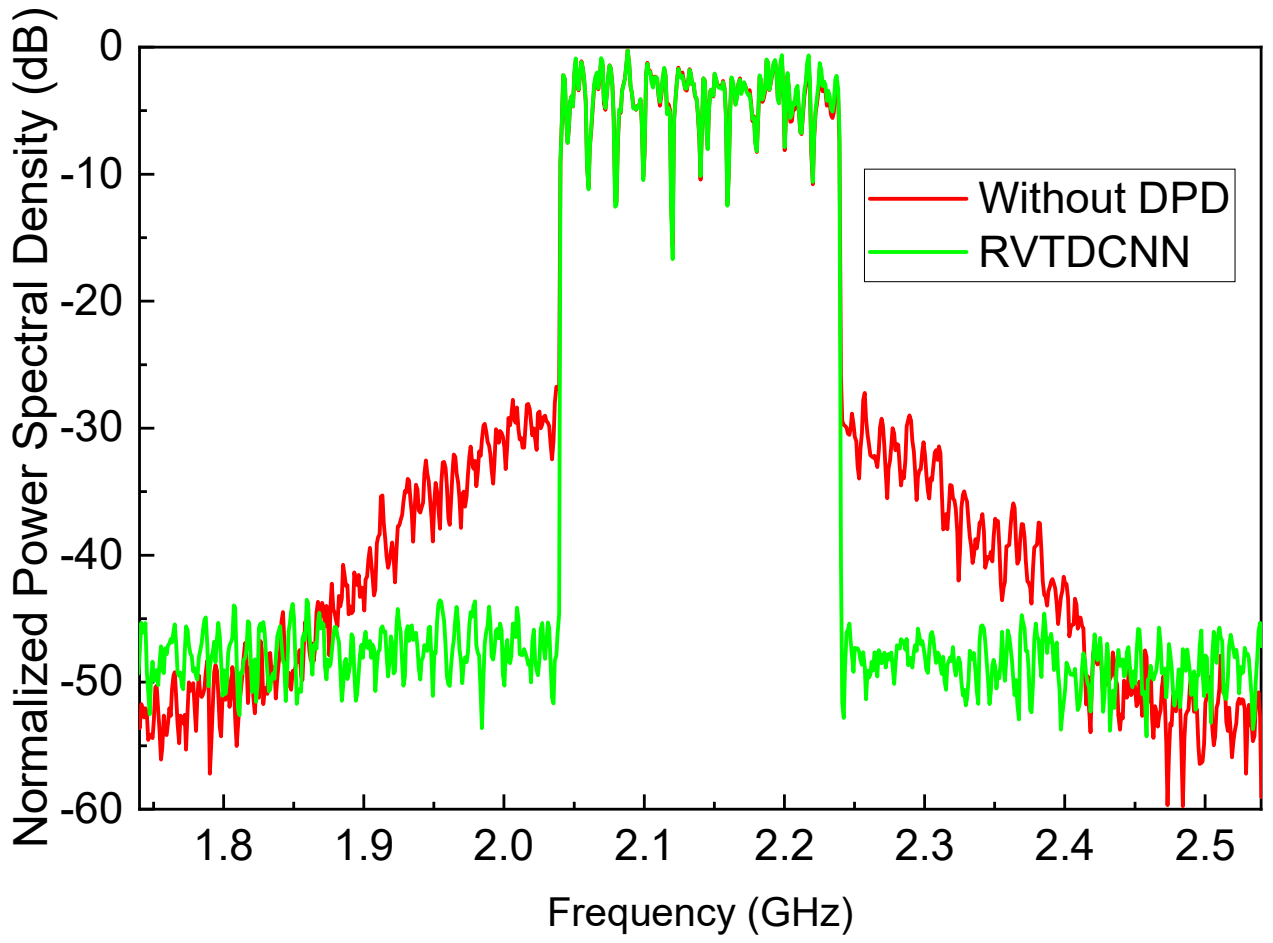

Fig. 13. Linearization performance of the PA using RVTDCNN at 200MHz OFDM source signal.

Fig. 13 shows the output spectrum after the linearization of the PA using the RVTDCNN model at the 200MHz OFDM source signal. The results show that under a wide signal bandwidth, the RVTDCNN model still has a significant linearization effect on the power amplifier.

## VII. CONCLUSION

In this paper, RVTDCNN is proposed for modeling the nonlinear and memory effects of wideband PA. RVTDCNN extracts the effective features of the two-dimensional input graph data with a convolutional structure. Doherty PA, with an OFDM signal from 40 MHz to 200 MHz, is tested to verify the effectiveness of the RVTDCNN model. For the PA with 100 MHz under different cases, the NMSE can reach about -36 dB, with an ACPR around -46 dBc with DPD. The results show that the RVTDCNN still has a good modeling effect when there are I/Q imbalance and DC offset, which verifies that the proposed model has strong adaptability. Compared with the existing shallow NN and DNN in terms of the number of model coefficients and FLOPs, the proposed RVTDCNN is verified to reduce the number of model coefficients by better than 50% under different signal bandwidth.

## APPENDIX

### *Extraction Effect of Pre-Designed Filter on Basis Function*

Because the pre-designed filter can completely capture the basis function features required for modeling, the number of neurons in fully connected layer required is significantly reduced, thereby reducing the complexity of the model. To verify that the pre-designed filter can generate rich basis functions, we introduce a baseband PA model [2], which is expressed as follows.

$$y(n)=\sum_{k=0}^{K-1}a_k x(n)\left|x(n)\right|^k+\sum_{k=1}^{K-1}\sum_{q=1}^{Q-1}c_{kq}x(n)\left|x(n-q)\right|^k \quad \text{(A.1)}$$

where $K$ and $Q$ are the nonlinearity order and the lagging cross terms index, respectively. $a_k$ and $c_{kq}$ are the coefficients of the model. For simplicity, the memory terms in (A.1) have been omitted

To simplify the derivation process, the coefficients of the model including $a_k$'s and $c_{kq}$'s have been omitted in Eq. (A.1). Thus, Eq. (A.1) can be expanded to be as follows.

$$\begin{aligned}
&y(n)\approx\left(I_{in}(n)+jQ_{in}(n)\right)+\left(I_{in}(n)+jQ_{in}(n)\right)\left|x(n)\right|+\\
&\left(I_{in}(n)+jQ_{in}(n)\right)\left|x(n)\right|^2+...+\left(I_{in}(n)+jQ_{in}(n)\right)\left|x(n-1)\right|+\\
&\left(I_{in}(n)+jQ_{in}(n)\right)\left|x(n-1)\right|^2+\cdots+\\
&\left(I_{in}(n)+jQ_{in}(n)\right)\left|x(n-2)\right|+\left(I_{in}(n)+jQ_{in}(n)\right)\left|x(n-2)\right|^2+...\\
&\quad=\Big(I_{in}(n)+I_{in}(n)\left|x(n)\right|+I_{in}(n)\left|x(n)\right|^2+I_{in}(n)\left|x(n-1)\right|+\\
&I_{in}(n)\left|x(n-1)\right|^2+I_{in}(n)\left|x(n-2)\right|+I_{in}(n)\left|x(n-2)\right|^2\Big)+\\
&+j\Big(Q_{in}(n)+Q_{in}(n)\left|x(n)\right|+Q_{in}(n)\left|x(n)\right|^2+Q_{in}(n)\left|x(n-1)\right|+\\
&Q_{in}(n)\left|x(n-1)\right|^2+Q_{in}(n)\left|x(n-2)\right|+Q_{in}(n)\left|x(n-2)\right|^2\Big)+\textit{other terms}
\end{aligned}\quad \text{(A.2)}$$

If these items in Eq. (A.2) can be fitted by the pre-designed filter, the pre-designed filter can generate rich basis functions, and the modeling performance of the proposed model can reach the modeling performance of the model. In the pre-designed filter, the 3*3 convolution kernels are used to extract features from the input signal. In the convolution process, the stride is set to be 1 until the input tensors are all convolved. And the convolution process is shown in Fig. 4. The convolution output under different convolution steps represents the capture of different local features of the input. We take out the local input convoluted by the first step.

$$\begin{aligned}
h_{11}=&I_{in}(n)+Q_{in}(n)+\left|x(n)\right|+I_{in}(n-1)+Q_{in}(n-1)\\
&+\left|x(n-1)\right|+I_{in}(n-2)+Q_{in}(n-2)+\left|x(n-2)\right|+b
\end{aligned}\quad \text{(A.3)}$$

where $b$ is the bias. For the convenience of calculation, the corresponding coefficient is ignored in the formula.

This paper deduced the item in Eq. (A.2) through the above local input, and the other input term with memory effect can be deduced through the same method. The convolution output described above is input into a nonlinear activation function to obtain the output of the pre-designed filter. The output $u_{11}$ can be expressed as follows.

$$\begin{aligned}
&u_{11}=\tanh\left(h_{11}\right)\\
&=\tanh\Big(I_{in}(n)+Q_{in}(n)+\left|x(n)\right|+I_{in}(n-1)+Q_{in}(n-1)+\\
&\left|x(n-1)\right|+I_{in}(n-2)+Q_{in}(n-2)+\left|x(n-2)\right|+b\Big)
\end{aligned}\quad \text{(A.4)}$$

where $\tanh(\cdot)$ is the activation function given.

The hyperbolic tangent sigmoid function can be expanded by the Taylor series, and the approximate output is described as follows.

$$\tanh(x)=x-\frac{1}{3}x^3+\frac{2}{15}x^5+-\cdots\left(\left|x\right|<\frac{\pi}{2}\right)\quad \text{(A.5)}$$

Bring Eq. (A.3) into $x^3$, while considering the critical correlation items related to the justification process, ignore the irrelevant, redundant terms. The result will be shown in Eq. (A.6).

$$\begin{aligned}
&h_{11}^3=\big(I_{in}(n)+Q_{in}(n)+\left|x(n)\right|+I_{in}(n-1)+Q_{in}(n-1)\\
&\quad+\left|x(n-1)\right|)+I_{in}(n-2)+Q_{in}(n-2)+\left|x(n-2)\right|+b\big)^3\\
&=6bI_{in}(n)\left|x(n)\right|+3I_{in}(n)\left|x(n)\right|^2+6bI_{in}(n)\left|x(n-1)\right|\\
&+3I_{in}(n)\left|x(n-1)\right|^2+6bI_{in}(n)\left|x(n-2)\right|+3I_{in}(n)\left|x(n-2)\right|^2+\\
&6bQ_{in}(n)\left|x(n)\right|+3Q_{in}(n)\left|x(n)\right|^2+6bQ_{in}(n)\left|x(n-1)\right|+3Q_{in}(n)\left|x(n-1)\right|^2\\
&+6bQ_{in}(n)\left|x(n-2)\right|+3Q_{in}(n)\left|x(n-2)\right|^2+\textit{other terms}
\end{aligned}\quad \text{(A.6)}$$

In summary, we combine Eq. (A.4) - Eq. (A.6), while omitting both the irrelevant, redundant terms and the multiplication factors. The output of the pre-designed filter can be expressed approximately using by Eq. (A.7).

$$\begin{aligned}
&u_{11}=\tanh\left(h_{11}\right)=h_{11}+h_{11}^3+\textit{other terms}\\
&\quad=\big(I_{in}(n)+Q_{in}(n)+\left|x(n)\right|+I_{in}(n-1)+Q_{in}(n-1)\\
&+\left|x(n-1)\right|)+I_{in}(n-2)+Q_{in}(n-2)+\left|x(n-2)\right|)+b\big)\\
&+\Big(I_{in}(n)\left|x(n)\right|+I_{in}(n)\left|x(n)\right|^2+I_{in}(n)\left|x(n-1)\right|+I_{in}(n)\left|x(n-1)\right|^2\\
&+I_{in}(n)\left|x(n-2)\right|+I_{in}(n)\left|x(n-2)\right|^2+Q_{in}(n)\left|x(n)\right|+\\
&Q_{in}(n)\left|x(n)\right|^2+Q_{in}(n)\left|x(n-1)\right|+Q_{in}(n)\left|x(n-1)\right|^2+\\
&Q_{in}(n)\left|x(n-2)\right|+Q_{in}(n)\left|x(n-2)\right|^2+\textit{other terms}\Big)
\end{aligned}\quad \text{(A.7)}$$

By arranging Eq. (A.7), Eq. (A.7) can be rewritten as follows.

$$\begin{aligned}
u_{11}=&\Big(I_{in}(n)+I_{in}(n)\left|x(n)\right|+I_{in}(n)\left|x(n)\right|^2+I_{in}(n)\left|x(n-1)\right|+\\
&I_{in}(n)\left|x(n-1)\right|^2+I_{in}(n)\left|x(n-2)\right|+I_{in}(n)\left|x(n-2)\right|^2\Big)+\\
&\Big(Q_{in}(n)+Q_{in}(n)\left|x(n)\right|+Q_{in}(n)\left|x(n)\right|^2+Q_{in}(n)\left|x(n-1)\right|+\\
&Q_{in}(n)\left|x(n-1)\right|^2+Q_{in}(n)\left|x(n-2)\right|+Q_{in}(n)\left|x(n-2)\right|^2\Big)+\textit{other terms}
\end{aligned}\quad \text{(A.8)}$$

Comparing Eq. (A.8) and Eq. (A.2), we can find that the linear term, the nonlinear term, and the lagging cross-terms in Eq. (A.2) both exist in Eq. (A.8). Therefore, the terms produced by the pre-designed filter are corresponding to the terms of the polynomials, and more terms can be provided. Namely, the pre-designed filter can generate enough rich basis set to get an excellent performance.

## REFERENCES

[1] V. Camarchia, M. Pirola, R. Quaglia, S. Jee, Y. Cho, and B. Kim, "The Doherty power amplifier: Review of recent solutions and trends," *IEEE Trans. Microw. Theory Techn.*, vol. 63, no. 2, pp. 559–571, Feb. 2015.

[2] Y. Liu, W. Pan, S. Shao, and Y. Tang, "A general digital predistortion architecture using constrained feedback bandwidth for wideband power amplifiers," *IEEE Trans. Microw. Theory Techn.*, vol. 63, no. 5, pp. 1544–1555, May 2015.

[3] A. Katz, J. Wood, and D. Chokola, "The evolution of PA linearization: From classic feedforward and feedback through analog and digital predistortion," *IEEE Microw. Mag.*, vol. 17, no. 2, pp. 32–40, Feb. 2016.

[4] Z. Popovic, "Amping up the PA for 5G: Efficient GaN power amplifiers with dynamic supplies," *IEEE Microw. Mag.*, vol. 18, no. 3, pp. 137–149, May 2017.

[5] X. Hu, T. Liu, Z. J. Liu, W. D. Wang, and F. M. Ghannouchi, "A Novel Single Feedback Architecture with Time-Interleaved Sampling for Multi-Band DPD," *IEEE Commun. Lett.*, vol. 23, no. 6, pp. 1033–1036, Jun. 2019.

[6] J. Reina-Tosina, M. Allegue-Martínez, C. Crespo-Cadenas, C. Yu, and S. Cruces, "Behavioral modeling and predistortion of power amplifiers under sparsity hypothesis," *IEEE Trans. Microw. Theory Techn.*, vol. 63, no. 2, pp. 745–753, Feb. 2015.

[7] Q. Zhang, W. Chen, and Z. Feng, "Reduced cost digital predistortion only with in-phase feedback signal," *IEEE Microw. Wireless Compon. Lett.*, vol. 28, no. 3, pp.257–259, Mar. 2018.

[8] J. Joung, C. K. Ho, K. Adachi, and S. Sun, "A survey on poweramplifier-centric techniques for spectrum-and energy-efficient wireless communications," *IEEE Commun. Surv. Tut.*, vol. 17, no. 1, pp. 315–333, Jan.–Mar. 2015.

[9] A. Cheaito, M. Crussière, J. Hélard, and Y. Louët, "Quantifying the Memory Effects of Power Amplifiers: EVM Closed-Form Derivations of Multicarrier Signals," *IEEE Wireless Commun. Lett.*, vol. 6, no. 1, pp. 34–37, Feb. 2017.

[10] S. Wang, M. Roger, and C. Lelandais-Perrault, "Impacts of crest factor reduction and digital predistortion on linearity and power efficiency of power amplifiers," *IEEE Trans. Circuits Syst. II, Exp. Briefs*, vol. 66, no. 3, pp. 407–411, Mar. 2019.

[11] J. Cai, C. Yu, L. Sun, S. Chen, and J. B. King, ''Dynamic behavioral modeling of RF power amplifier based on time-delay support vector regression,'' *IEEE Trans. Microw. Theory Techn.*, vol. 67, no. 2, pp. 533–543, Feb. 2019.

[12] R. N. Braithwaite, "Digital Predistortion of an RF Power Amplifier Using a Reduced Volterra Series Model with a Memory Polynomial Estimator," *IEEE Trans. Microw. Theory Techn.*, vol. 65, no. 10, pp. 3613–3623, Oct. 2017.

[13] A. R. Belabad and S. A. Motamedi, "A novel generalized parallel two-box structure for behavior modeling and digital predistortion of RF power amplifiers at LTE applications," *Circuits Syst. Signal Process.*, vol. 37, no. 7, pp. 2714–2735, Jul. 2018.

[14] R. Hongyo, Y. Egashira, T. M. Hone, and K. Yamaguchi, "Deep Neural Network-Based Digital Predistorter for Doherty Power Amplifiers," *IEEE Microw. Wireless Compon. Lett.*, vol. 29, no. 2, pp. 146–148, Feb. 2019

[15] T. Liu, S. Boumaiza, and F. M. Ghannouchi, "Dynamic behavioral modeling of 3G power amplifiers using real-valued time-delay neural networks," *IEEE Trans. Microw. Theory Techn.*, vol. 52, no. 3, pp. 1025–1033, Mar. 2004.

[16] M. Rawat, K. Rawat, and F. M. Ghannouchi, "Adaptive Digital Predistortion of Wireless Power Amplifiers/Transmitters Using Dynamic Real-Valued Focused Time-Delay Line Neural Networks," *IEEE Trans. Microw. Theory Techn.*, vol. 58, no. 1, pp. 95-104, Jan. 2010.

[17] M. Rawat and F. M. Ghannouchi, "A Mutual Distortion and Impairment Compensator for Wideband Direct-Conversion Transmitters Using Neural Networks," *IEEE Trans. Broadcast.*, vol. 58, no. 2, pp. 168-177, Jun. 2012.

[18] M. Hui, T. Liu, M. Zhang, Y. Ye, D. Shen, and X. Ying, "Augmented radial basis function neural network predistorter for linearisation of wideband power amplifiers," *Electron. Lett.*, vol. 50, no. 12, pp. 877–879, Jun. 2014.

[19] D. Wang, M. Aziz, M. Helaoui, and F. M. Ghannouchi, "Augmented real-valued time-delay neural network for compensation of distortions and impairments in wireless transmitters," *IEEE Trans. Neural Netw. Learn. Syst.*, vol. 30, no. 1, pp. 242–254, Jun. 2019.

[20] X. You, C. Zhang, X. Tan, S. Jin, and H. Wu, "Ai for 5G: Research directions and paradigms," *Sci. China Inf. Sci.*, vol. 62, no. 2, p. 21301, Feb. 2019.

[21] Z. M. Fadlullah, B. Mao, F. Tang, and N. Kato, "Value iteration architecture based deep learning for intelligent routing exploiting heterogeneous computing platforms," *IEEE Trans. Comput.*, vol. 68, no. 6, p. 939–950, Jun 2019.

[22] Y. Wang, M. Liu, J. Yang, and G. Gui, "Data-driven deep learning for automatic modulation recognition in cognitive radios," *IEEE Trans. Veh. Tech.*, vol. 68, no. 4, pp. 4074–4077, Apr. 2019.

[23] Z. M. Fadlullah, F. Tang, B. Mao, N. Kato, O. Akashi, T. Inoue, and K. Mizutani, "State-of-the-art deep learning: Evolving machine intelligence toward tomorrow's intelligent network traffic control systems," *IEEE Commu. Surv. Tut.*, vol. 19, no. 4, pp. 2432–2455, 2017.

[24] N. Kato, Z. M. Fadlullah, B. Mao, F. Tang, O. Akashi, T. Inoue, and K. Mizutani, "The deep learning vision for heterogeneous network traffic control: Proposal, challenges, and future perspective," *IEEE Wireless Commun.*, vol. 24, no. 3, pp. 146–153, 2016.

[25] J. Schmidhuber, "Deep learning in neural networks: An overview," *Neural netw.*, vol. 61, pp. 85–117, 2015.

[26] W. Liu, Z. Wang, X. Liu, N. Zeng, Y. Liu, and F. E. Alsaadi, "A survey of deep neural network architectures and their applications," *Neurocomputing*, vol. 234, pp. 11–26, Apr. 2017.

[27] M. Robnik-Šikonja, "Data generators for learning systems based on RBF networks," *IEEE Trans. Neural Netw. Learn. Syst.*, vol. 27, no. 5, pp. 926–938, May 2016.

[28] Z. Liu, X. Hu, T. Liu, X. Li, W. Wang, and F. M. Ghannouchi, "Attention-Based Deep Neural Network Behavioral Model for Wideband Wireless Power Amplifiers," *IEEE Microw. Wireless Compon. Lett.*, vol. 30, no. 1, pp. 82–85, Jan. 2020.

[29] Y. LeCun, L. Bottou, Y. Bengio, P. Haffner, "Gradient-based learning applied to document recognition," *Proc. IEEE*, vol. 86, no. 11, pp. 2278–2324, 1998.

[30] C. Szegedy, W. Liu, Y. Jia, P. Sermanet, S. Reed, D. Anguelov, D. Erhan, V. Vanhoucke, and A. Rabinovich, "Going deeper with convolutions," in *Proc. IEEE Conf. Comput. Vis. Pattern Recognit.*, 2015, pp. 1–9.

[31] A. Graves, A.-r. Mohamed, and G. Hinton, "Speech recognition with deep recurrent neural networks," in 2013 *IEEE int. Conf. Acoust. Speech Signal Process.*, 2013, pp. 6645–6649.

[32] M. Schuster and K. K. Paliwal, "Bidirectional recurrent neural networks," *IEEE Trans. Signal Process.*, vol. 45, no. 11, pp. 2673–2681, 1997.

[33] G. Gui, H. Huang, Y. Song, and H. Sari, "Deep learning for an effective non-orthogonal multiple access scheme," *IEEE Trans. Veh. Technol.*, vol. 67, no. 9, pp. 8440–8450, 2018.

[34] H. Huang, Y. Song, J. Yang, G. Gui, and F. Adachi, "Deep-learningbased millimeter-wave massive mimo for hybrid precoding," *IEEE Trans. Veh. Technol.*, vol. 68, no. 3, pp. 3027–3032, 2019.

[35] F. Meng, P. Chen, L. Wu, and X. Wang, "Automatic modulation classification: A deep learning enabled approach," *IEEE Trans. Veh. Technol.*, vol. 67, no. 11, pp. 10760–10772, 2018.

[36] F. Tang, B. Mao, Z. M. Fadlullah, and N. Kato, "On a novel deep learning-based intelligent partially overlapping channel assignment in sdn-iot," *IEEE Commun. Mag.*, vol. 56, no. 9, pp. 80–86, 2018.

[37] Z. M. Fadlullah, F. Tang, B. Mao, J. Liu, and N. Kato, "On intelligent traffic control for large-scale heterogeneous networks: A value matrixbased deep learning approach," *IEEE Commun. Lett.*, vol. 22, no. 12, pp. 2479–2482, 2018.

[38] M. Liu, T. Song, J. Hu, J. Yang, and G. Gui, "Deep learning-inspired message passing algorithm for efficient resource allocation in cognitive radio networks," *IEEE Trans. Veh. Technol.*, vol. 68, no. 1, pp. 641–653, 2019.

[39] P. Chen, S. Alsahali, A. Alt, J. Lees, and P. J. Tasker, "Behavioral Modeling of GaN Power Amplifiers Using Long Short-Term Memory Networks," *2018 International Workshop on Integrated Nonlinear Microwave and Millimetre-wave Circuits (INMMIC)*, Brive La Gaillarde, 2018, pp. 1-3.

[40] J. Sun, W. Shi, Z. Yang, J. Yang, and G. Gui, "Behavioral Modeling and Linearization of Wideband RF Power Amplifiers Using BiLSTM Networks for 5G Wireless Systems," *IEEE Trans. Veh. Technol.*, Jun. 2019, in Press, doi: 10.1109/TVT.2019.2925562.

[41] D. Phartiyal and M. Rawat, "LSTM-Deep Neural Networks based Predistortion Linearizer for High Power Amplifiers," *2019 National Conference on Communications (NCC)*, Bangalore, India, 2019, pp. 1-5.

[42] T. J. Liu, Y. Ye, S. Y. Yin, H. Chen, G. M. Xu, Y. L. Lu, and Y. Chen, "Digital Predistortion Linearization with Deep Neural Networks for 5G Power Amplifiers," *2019 European Microwave Conference in Central Europe (EuMCE)*, May 2019, pp. 216–219.

[43] F. Mkadem and S. Boumaiza, "Physically inspired neural network model for RF power amplifier behavioral modeling and digital predistortion," *IEEE Trans. Microw. Theory Technol.*, vol. 59, no. 4, pp. 913–923, Apr. 2011.

[44] D. P. Kingma and J. Ba, "Adam: A Method for Stochastic Optimization," in *Proc. 3rd International Conference for Learning Representations (ICLR 2015)*, San Diego, USA, 2015, pp. 7-9.

[45] S. Haykin, *Neural Networks: A Comprehensive Foundation*. Upper Saddle River, NJ, USA: Prentice-Hall, 1999.

[46] D. R. Morgan, Z. Ma, J. Kim, M. G. Zierdt, and J. Pastalan, "A generalized memory polynomial model for digital predistortion of RF power amplifiers," *IEEE Trans. Signal Process.*, vol. 54, no. 10, pp. 3852–3860, Oct. 2006.

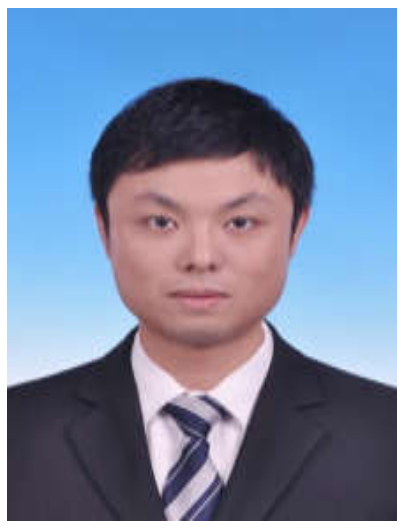

**Xin Hu** (M'18) earned the B.Sc. and Ph.D. degrees in electrical information engineering from Huazhong University of Science and Technology , Wuhan, China, and Institute of Electrics, Chinese Academy of Sciences, Beijing, China, in 2007 and 2012, respectively. From 2012 to 2016, he was a senior engineer in Aerospace Science and Technology Corporation. In 2016, he joined the Electronics Technology Lab at Beijing University of Posts and Telecommunications, where he is currently an Associate Professor. From 2019 to 2020, he was a Visiting Scholar with the iRadio Laboratory at the University of Calgary, Calgary, AB, Canada. From 2019, he is an Adjunct Researcher at the University of Calgary, Calgary, AB, Canada. His current research interests include digital predistortion of nonlinear power amplifiers and the application of signal processing techniques to RF and microwave problems, dynamic wireless resource management. He received the National Natural Science Foundation of China in 2018. He was involved in several projects funded by the National High Technology Research and Development Program of China and the National Natural Science Foundation of China.

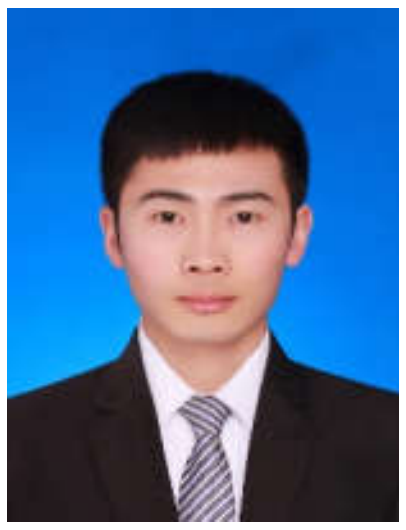

**Zhijun Liu** He received the B.Sc. degree in Electronics Science and Technology from Nanjing University of Posts and Telecommunications, China, in 2017, and he is currently pursuing the Ph.D. degrees in Electronics Science and Technology from Beijing University of Posts and Telecommunications, China. His research interests include digital predistortion of nonlinear power amplifiers and the application of deep learning techniques to RF and microwave problems.

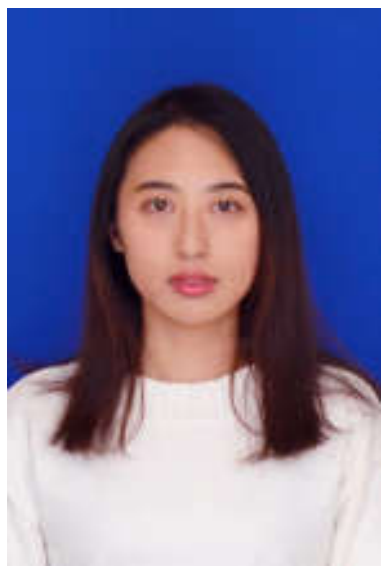

**Xiaofei Yu** She received the B.Sc. degree in Electronics Science and Technology from Beijing University of Posts and Telecommunications, China, in 2018, and she is currently pursuing the M.Sc degrees in Electronics Science and Technology from Beijing University of Posts and Telecommunications, China. Her research interests include digital predistortion of nonlinear power amplifiers and the application of deep learning techniques to RF and the crowd sensing.

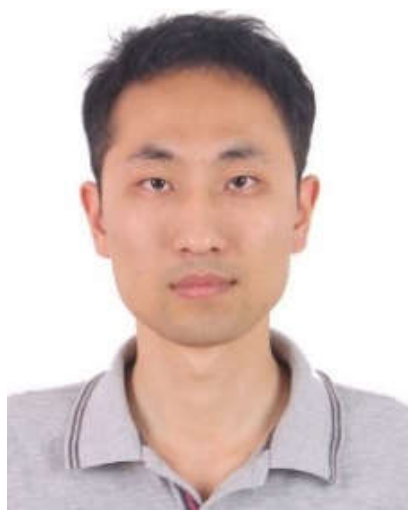

**Yulong Zhao** (S'16) was born in Henan, China, in 1983. He received the B.S. degree in applied physics and M.S. degree in electronic engineering from Xidian University, Xi'an, China, in 2006 and 2009, respectively. He is currently pursuing the Ph.D. degree at the University of Calgary, Calgary, AB, Canada. From 2009 to 2015, he was a Microwave Engineer and Project Leader with the ZTE Corporation. He was with the Radio Remote Unit Development Department, where he designed high-power amplifiers for 3G and 4G wireless communications. He is currently with the Intelligent RF Radio Technology Laboratory, University of Calgary. His current research interests include high-efficiency and wideband RF PAs, MMIC PAs, microwave passive components and digital pre-distortion.

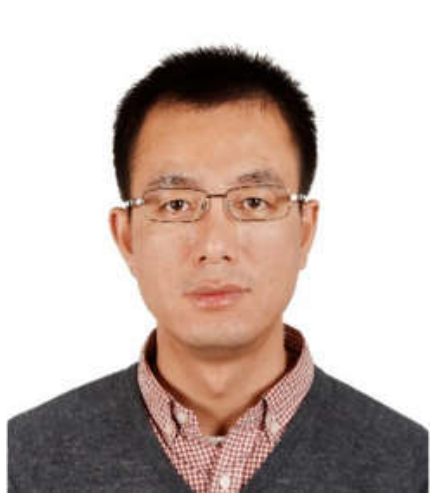

**Wenhua Chen** (S'03–M'07–SM'11) received the B.S. degree in microwave engineering from the University of Electronic Science and Technology of China (UESTC), Chengdu, China, in 2001, and the Ph.D. degree in electronic engineering from Tsinghua University, Beijing, China, in 2006. From 2010 to 2011, he was a Post-Doctoral Fellow with the Intelligent RF Radio Laboratory(iRadio Lab), University of Calgary, Calgary, AB, Canada. He is currently an Associate Professor with the Department of Electronic Engineering, Tsinghua University. He has authored or coauthored more than 200 journals and conference papers. His current research interests include energy-efficient power amplifier (PA) design and linearization, millimeter wave and terahertz integrated circuit and system. Dr. Chen was a recipient of the 2015 Outstanding Youth Science Foundation of NSFC, the 2014 URSI Young Scientist Award, and the Student Paper Awards of several international conferences. He is an Associate Editor of the IEEE Transactions on Microwave Theory and Techniques.

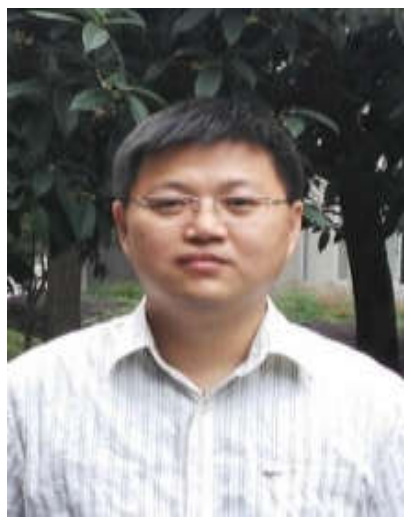

**Biao Hu** received the B.S. degree in applied physics, the M.S. degree and Ph.D. degree in physical electronics from University of Electronic Science and Technology of China, Chengdu, China, in 2005, 2008 and 2014, respectively. From 2016, he is an Associate Professor in the School of Electronic Science and Engineering, University of Electronic Science and Technology of China. From 2019, he is a Visiting Scholar with the iRadio Laboratory at the University of Calgary, Calgary, AB, Canada. His current research interests include microwave wireless power transmission, microwave/millimeter wave integrated circuits and antennas, high power microwave technique and its applications.

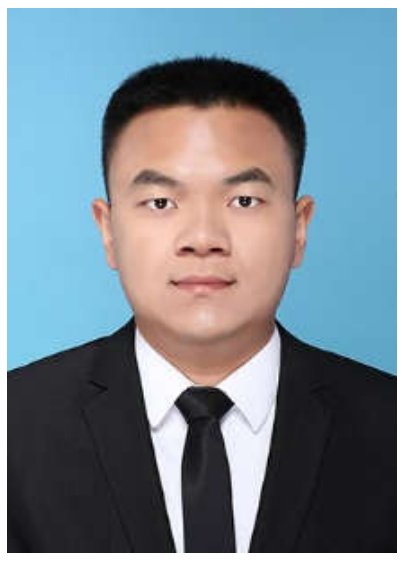

**Xuekun Du** (S'14–M'19) received the M.Sc. degree in Communication & Information System from Liaoning Technical University in 2014 and the Ph.D. degree in Communication & Information Systems from the University of Electronic Science and Technology of China in 2019. He is currently a Postdoctoral Fellow with the Intelligent RF Radio Laboratory, Department of electrical and computer engineering, University of Calgary, Calgary, AB, Canada. From 2016 to 2018, he has been a visiting Ph.D. Student with the Intelligent RF Radio Laboratory, Department of electrical and computer engineering, University of Calgary, Calgary, AB, Canada. His current research interests include passive circuit design, high efficient wideband power amplifier design, MMIC PA design, active devices modeling, artificial neural network modeling, measurements and characterization techniques.

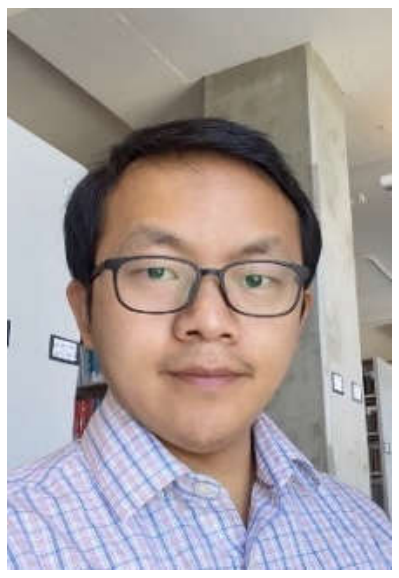

**Xiang Li** (S'13-M'17) received the B.Sc. and M.Sc. degrees in Electronic Engineering from Tsinghua University, Beijing, China, in 2009 and 2012, respectively. He earned the Ph.D. degree in electrical and computer engineering from University of Calgary in 2018. He is currently working as post-doctoral fellow at University of Calgary, Calgary, AB, Canada. His research interests include multi-band/wide-band power amplifier design, outphasing transmitter design and MMIC power amplifiers for wireless and satellite communication. Mr. Li was the recipient of the Student Paper Award of 2010 Asia-Pacific Microwave Conference (APMC).

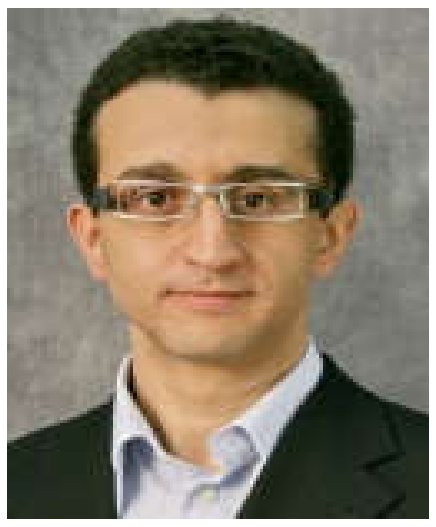

**Mohamed Helaoui** (S'06, M'09) received the M.Sc. degree in communications and information technology from École Supérieure des Communications de Tunis, Tunisia, in 2003 and the Ph.D. degree in electrical engineering from the University of Calgary in 2008. He is currently an associate professor at the department of electrical and computer engineering of the University of Calgary. His current research interests include digital signal processing, power efficiency enhancement for wireless transmitters, switching mode power amplifiers, and advanced transceiver design for software defined radio and millimeter-wave applications. His research activities have led to over 60 publications and 7 patents (pending). He is a member of the COMMTTAP chapter in the IEEE southern Alberta section.

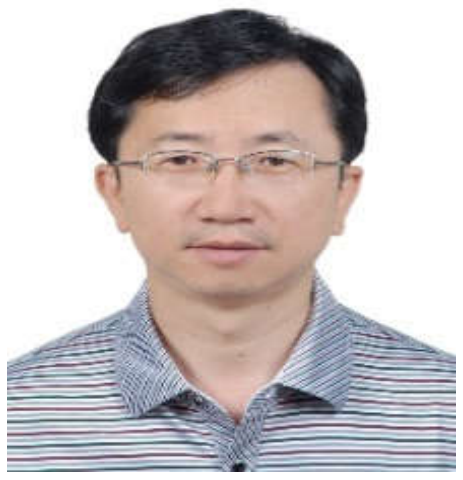

**Weidong Wang** He received the Ph.D. degree from Beijing University of Posts and Telecommunications in 2002. He is now the Professor of School of Electronic Engineering at Beijing University of Posts and Telecommunications. He takes the role of expert of National Natural Science Foundation and member of China Association of Communication. His research interests include communication system, radio resource management, Internet of things and signal processing.

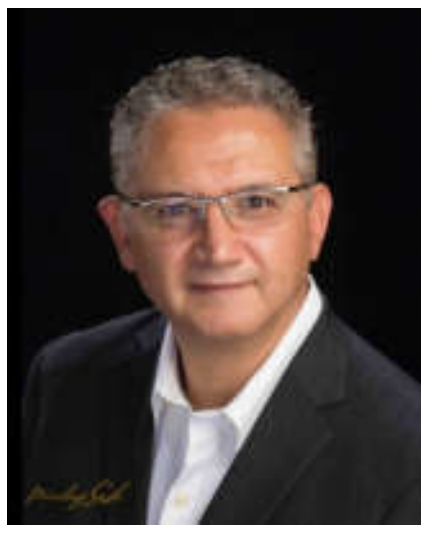

**Fadhel Ghannouchi** is currently a professor, Alberta Innovates/Canada Research Chair and Director of the iRadio Laboratory in the Department of Electrical and Computer Engineering at the University of Calgary, Alberta, Canada and a part-time Thousand Talent Professor at Department of Electronics Engineering, Tsinghua University, Beijing China. His research interests are in the areas of RF and wireless communications, nonlinear modeling of microwave devices and communications systems, design of power- and spectrum-efficient microwave amplification systems and design of SDR systems for wireless, optical and satellite communications applications. He is Fellow of the Academy of Science of the Royal Society of Canada, Fellow of the Canadian Academy of Engineering, Fellow of the Engineering Institute of Canada, Fellow of the Institution of Engineering and Technology (IET) and Fellow of the Institute of Electrical and Electronic Engineering (IEEE). He published more than 800 referred papers, 6 books and hold 25 patents (3 pending). Prof. Ghannouchi is the co-founder of three university spun-off companies.